\documentclass[sigconf]{acmart}
\AtBeginDocument{%
  }

\copyrightyear{2025}
\acmYear{2025}
\acmDOI{10.1145/3746027.3755410}
\setcopyright{acmlicensed}\acmConference[MM '25]{Proceedings of the 33rd ACM International Conference on Multimedia}{October 27--31, 2025}{Dublin, Ireland}
\acmBooktitle{Proceedings of the 33rd ACM International Conference on Multimedia (MM '25), October 27--31, 2025, Dublin, Ireland}
\acmISBN{979-8-4007-2035-2/2025/10}

\usepackage{enumitem}
\usepackage{subfig}
\usepackage{bm}
\usepackage{threeparttable}
\begin{document}

\title{I$^3$-MRec: Invariant Learning with Information Bottleneck for Incomplete Modality Recommendation}


\author{Huilin Chen}
\affiliation{%
  \institution{Hefei University of Technology}
  \city{Anhui}
  \country{China}
  }
\email{ClownClumsy@outlook.com}

\author{Miaomiao Cai}
\affiliation{%
  \institution{Hefei University of Technology}
  \city{Anhui}
  \country{China}
  }
\email{cmm.hfut@gmail.com}

\author{Fan Liu}
\authornotemark[1]
\affiliation{%
  \institution{National University of Singapore}
  \city{Singapore}
  \country{Singapore}
  }
\email{liufancs@gmail.com}

\author{Zhiyong Cheng}
\authornote{Corresponding author.}
\affiliation{%
  \institution{Hefei University of Technology}
  \city{Anhui}
  \country{China}
  }
\email{jason.zy.cheng@gmail.com}

\author{Richang Hong}
\affiliation{%
  \institution{Hefei University of Technology}
  \city{Anhui}
  \country{China}
  }
\email{hongrc.hfut@gmail.com}

\author{Meng Wang}
\affiliation{%
  \institution{Hefei University of Technology}
  \city{Anhui}
  \country{China}
  }
\email{eric.mengwang@gmail.com}

\renewcommand{\shortauthors}{Huilin Chen et al.}
\begin{abstract}
Multimodal recommender systems (MRS) improve recommendation performance by integrating complementary semantic information from multiple modalities. However, the assumption of complete multimodality rarely holds in practice due to missing images and incomplete descriptions, hindering model robustness and generalization. To address these challenges, we introduce a novel method called \textbf{I$^3$-MRec}, which uses \textbf{I}nvariant learning with \textbf{I}nformation bottleneck principle for \textbf{I}ncomplete \textbf{M}odality \textbf{Rec}ommendation. To achieve robust performance in missing modality scenarios, I$^3$-MRec enforces two pivotal properties: (i) cross-modal preference invariance, ensuring consistent user preference modeling across varying modality environments, and (ii) compact yet effective multimodal representation, as modality information becomes unreliable in such scenarios, reducing the dependence on modality-specific information is particularly important. By treating each modality as a distinct semantic environment, I$^3$-MRec employs invariant risk minimization (IRM) to learn preference-oriented representations.  In parallel, a missing-aware fusion module is developed to explicitly simulate modality-missing scenarios. Built upon the Information Bottleneck (IB) principle, the module aims to preserve essential user preference signals across these scenarios while effectively compressing modality-specific information. Extensive experiments conducted on three real-world datasets demonstrate that I$^3$-MRec consistently outperforms existing state-of-the-art MRS methods across various modality-missing scenarios, highlighting its effectiveness and robustness in practical applications. 
\end{abstract}



\maketitle

\begin{figure}[t]
    \vspace{0.0cm}
    \centering
    \includegraphics[width=0.9\linewidth]{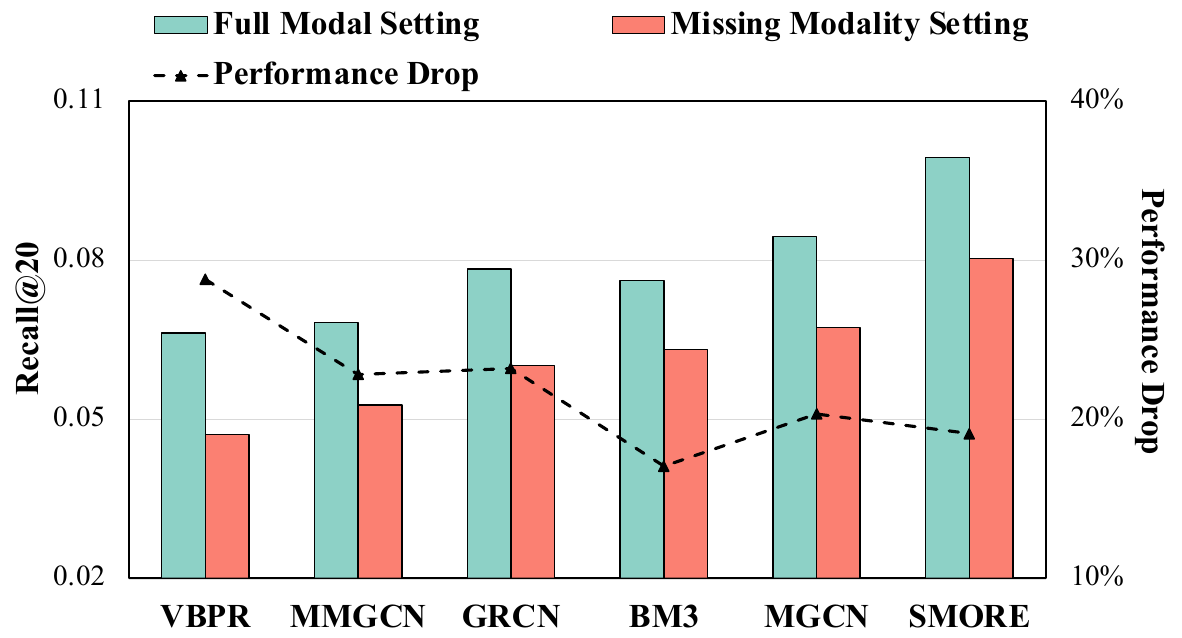}
    \caption{Performance of MRS methods on the Baby dataset under two settings. “Full Modality” indicates no missing modality during training and testing. “Missing Modality” follows the MTMT setup (Section 4.3), with random modality missingness in both phases.}
    \label{fig:performance_drop}
    \vspace{-0.3 cm}
\end{figure}

\section{Introduction}
Multimodal data has experienced explosive growth across various online platforms, including Amazon~\footnote{https://www.amazon.com}, YouTube\footnote{https://www.youtube.com}, and TikTok\footnote{https://www.tiktok.com}. To mitigate information overload and help users discover relevant items, Multimodal Recommender Systems (MRS)~\cite{liu2023semantic, BM32020Arxiv,wang2023generative,Wei2023LightGT,yu2023multi,lin2024gume,xu2024mentor,lin2025contrastive,ADRL2024MM} leverage rich information from various modalities (e.g., item images, review texts) to model user preferences and generate more accurate recommendations. However, most existing MRS methods assume the availability of complete modality information to effectively extract recommendation-relevant signals~\cite{bai2024multimodality, malitesta2024we}. This assumption often fails in real-world settings, where incomplete modality data is often encountered in real-world applications, such as information retrieval~\cite{wang2018lrmm, wu2020joint, malitesta2024dealing}, computer vision~\cite{wang2023multi}, and video-sharing platforms~\cite{li2025generating}. As illustrated in Fig.~\ref {fig:performance_drop}, many existing multimodal recommendation models suffer from performance degradation or even fail when confronted with missing modalities\footnote{For fair comparison, results are reproduced using the official code of the referenced method on the same dataset. }. This frequent occurrence poses a substantial challenge to the robustness and practical applicability of MRS in real-world deployments.

To address the challenge of missing modalities, several efforts have been proposed~\cite{parthasarathy2020training, wang2020multimodal,wang2018lrmm, wu2020joint, bai2024multimodality, ganhor2024multimodal, lin2023contrastive, li2025generating}. Early approaches often resorted to discarding items with incomplete modality information~\cite{parthasarathy2020training, wang2020multimodal}, but this strategy worsens data sparsity and leads to substantial performance degradation. More recent efforts have attempted to recover missing modality data through various content generation techniques~\cite{wang2018lrmm, wu2020joint, bai2024multimodality, ganhor2024multimodal, lin2023contrastive, li2025generating}. For instance, CI$^2$MG~\cite{lin2023contrastive} employs hypergraph convolution and cross-modal transport to generate missing modality features. In addition, representation learning approaches, such as feature propagation~\cite{malitesta2024dealing} and invariant learning~\cite{bai2024multimodality}, have been proposed to learn robust preference representations that remain effective under missing modality scenarios.

Although these methods have demonstrated promising effectiveness, the core challenges associated with missing modalities remain inadequately addressed:  \textit{How should an optimal recommender system perform when confronted with modality missingness?} On the one hand, generative methods~\cite{lin2023contrastive, li2025generating} often require extensive and time-consuming pre-training for each modality. This requirement contradicts the missing modality assumption, as it presumes the availability of large-scale, fully annotated multimodal data. 
On the other hand, existing representation-based approaches overlook the distinct characteristics of each modality. For example, MILK~\cite{bai2024multimodality} adopts a modality-wise mixup strategy to simulate modality-missing environments. However, such operations can disrupt modality-specific semantic structures, ultimately making the model more vulnerable to performance degradation under real-world missing modality conditions. 

We argue that a robust recommendation model should satisfy two key properties:
\begin{itemize}[leftmargin=0.35cm,topsep=3pt]
\item \textbf{Cross-modal Preference Invariance:} User preferences are inherently stable across different modality sources~\cite{du2022invariant}. A robust model should therefore learn preference-oriented user/item representations that remain invariant across modalities. This ensures that each individual modality representation can sufficiently support accurate preference prediction, thereby mitigating the negative impact of missing modalities.

\item \textbf{Compact yet Effective Multimodal Representation:} The learned multimodal representations should maximally preserve user preference information pertinent to recommendation tasks while minimizing dependence on the raw modality information. As modality information becomes unreliable in such scenarios, especially when missing modalities randomly occur during training, reducing the dependence on modality-specific information is particularly important~\cite{ma2021smil,woo2023towards, saeed2024modality}. 

\end{itemize}

These two principles together promote recommendation robustness by ensuring robust representations and reducing the model’s dependency on potentially unavailable modality data. To address this critical challenge, we propose \textbf{I$^3$-MRec} (Invariant learning with Information Bottleneck for Incomplete-Modality Recommendation), a unified framework specifically designed for robust recommendation under modality-missing scenarios. To achieve the goal, I$^3$-MRec adopts an invariant learning framework that explicitly treats each modality as a distinct semantic environment. It leverages a GCN-based architecture combined with Invariant Risk Minimization (IRM) to learn robust user and item representations (\textit{thereby satisfying the first property}). Furthermore, we introduce a missing-aware fusion module guided by the Information Bottleneck (IB) principle~\cite{tishby2000information, tishby2015deep, hu2024survey} to explicitly simulate diverse modality-missing scenarios during training, maximizing mutual information between multimodal representations across these scenarios while minimizing the mutual information between the multimodal representation and each raw-modality feature. (\textit{in alignment with the second property}). We have released the code and relevant parameter settings to facilitate repeatability as well as further research\footnote{https://github.com/HuilinChenJN/I3-MRec.}.

In summary, the main contributions of this work are as follows:

\begin{itemize}[leftmargin=0.35cm,topsep=3pt]

\item We propose I$^3$-MRec, a novel framework that addresses the limitations of existing multimodal recommender systems in modality-missing scenarios. By integrating invariant learning with the information bottleneck principle, our method enhances model robustness and generalization.

\item We introduce a principled invariant learning approach that models each modality as a distinct semantic environment. This enables learning of preference-oriented user/item representations, improving resilience to missing modality inputs.

\item We design a missing-aware fusion module guided by the Information Bottleneck principle, which selectively retains preference-relevant information while compressing modality-specific information, resulting in compact yet effective multimodal representations.

\item We have conducted extensive experiments on three real-world datasets, demonstrating the superiority of our method over state-of-the-art baselines.
\end{itemize}
\section{Related Work}
\label{sec:relatedwork}
\subsection{Multimodal Recommendation and Modality Missingness }
Multimodal Recommender Systems leverage various multimedia content to enhance user-item interactions, offering significant advantages in applications such as online shopping~\cite{kang2017visually,liu2022disentangled}, video sharing platforms~\cite{wei2019mmgcn, cai2022adaptive}, and social networks~\cite{zhou2023enhancing}. Early approaches~\cite{chen2017attentive, he2016vbpr, kang2017visually} incorporated multimodal data as auxiliary features to enhance item representations. For instance, VBPR~\cite{he2016vbpr} integrates visual embeddings with item ID embeddings to better model user preferences. With the development of Graph Neural Networks (GNNs)\cite{He@LightGCN, wang2019ngcf, Liu2021IMP_GCN,Liu20205SALLM}, subsequent work extended these models to incorporate multimodal signals into graph-based learning\cite{wei2019mmgcn, Wei2020GRCN, ma2024multimodal, cai2021heterogeneous}, improving the quality of user and item node representations. Some methods, such as LATTICE~\cite{zhang2021mining}, capture item-item semantic affinities via modality-aware graphs. while DualGNN~\cite{wang2021dualgnn} models user-user relations to reveal latent preference patterns.  In addition, recent advances have incorporated attention mechanisms~\cite{liu2022disentangled}, self-supervised learning~\cite{tao2022self, BM32020Arxiv}, and contrastive learning~\cite{lin2023contrastive, wei2023multi} to better align and integrate cross-modality information.

Despite their effectiveness, these methods typically assume the availability of complete modality information. In practice, missing modalities can lead to significant performance degradation~\cite{malitesta2024we, bai2024multimodality}. To address this challenge, existing solutions fall into two broad categories: Generation-based approaches aim to reconstruct missing modality features through modality generation techniques~\cite{wang2018lrmm, wu2020joint, bai2024multimodality, ganhor2024multimodal, lin2023contrastive, li2025generating}. While effective in controlled settings, these methods typically incur high computational costs and lack flexibility when handling random or modality-specific missing patterns. e.g., MoDiCF~\cite{li2025generating} requires an auxiliary diffusion model to reconstruct missing modalities. Representation Learning-based methods focus on directly learning robust representations that remain effective even when specific modalities are missing~\cite{bai2024multimodality, malitesta2024dealing}. For instance, MILK~\cite{bai2024multimodality} introduces a modality-invariant learning strategy; however, it assumes complete modality availability during training, which limits its applicability in real-world scenarios. By formulating the task within an invariant learning framework and incorporating an IB-guided fusion strategy, I$^3$-MRec generalizes effectively across various modality-missing scenarios.

\subsection{Invariant Learning in Recommendation}
Invariant learning (IL)~\cite{rojas2018invariant, arjovsky2019invariant, ahuja2020invariant,creager2021environment, li2022learning} is emerging as a pivotal technique to enhance model generalization. This approach often posits that certain stable features exist within the data that causally determine the target labels. Moreover, the relationship between these stable features and labels remains invariant in different environments~\cite{you2020graph}. Recently, efforts have been made to incorporate invariant learning principles into recommendation systems~\cite{wang2022invariant, du2022invariant, zhang2023invariant, zhang2023reformulating, bai2024multimodality}. For example, InvPref~\cite{wang2022invariant} estimates heterogeneous environments corresponding to different types of latent bias, while InvRL~\cite{du2022invariant} exploits spurious correlations in user-item interactions caused by modality noise to differentiate environment sets. Departing from MILK~\cite{bai2024multimodality}, which ensures that modality-specific preferences remain stable in cyclic mixup-based heterogeneous modality environments, we treat modality-missing patterns as heterogeneous environments to learn stable, preference-oriented modality representations.

\subsection{Information Bottleneck}
The Information Bottleneck (IB) principle, based on information theory, is widely applied in machine learning tasks such as model robustness~\cite{wu2020graph, zhang2022improving, ding2023robust}, fairness~\cite{gronowski2023classification}, explainability~\cite{espinosa2022concept, wang2023empower}, and recommendation systems~\cite{wei2022contrastive, yang2024graph, yang2025less, zhao2025dvib}. For input data $X$, hidden representation 
$Z$, and prediction label $Y$, IB principle suggests that an effective representation retains minimal sufficient information for the prediction task~\cite{tishby2000information, tishby2015deep, hu2024survey}: $Max:I(Y;Z) - \beta I(X;Z)$. where $I(Y;Z)$ and $I(X;Z)$  are the mutual information between variables, with $\beta$ balancing the two terms. Calculating mutual information for continuous variables is difficult, especially in deep learning. Approximations using neural networks, such as MINE~\cite{belghazi2018mutual}, InfoNCE~\cite{oord2018representation}, and variational methods~\cite{tishby2015deep}, are commonly used. Recently, CLUB~\cite{cheng2020club} has emerged to estimate the upper bound of mutual information using a log-ratio contrastive loss, which is more general for high-dimensional tasks without prior assumptions. 

This work applies IB learning to multimodal recommender systems, making them robust to modality missingness. Given the difficulty of estimating the upper bound of mutual information, we use InfoNCE and CLUB to approximate and optimize the mutual information between modality features.

\section {Preliminaries}
\label{sec:method}

\begin{figure*}[t]
    \centering
    \includegraphics[width=0.9\linewidth]{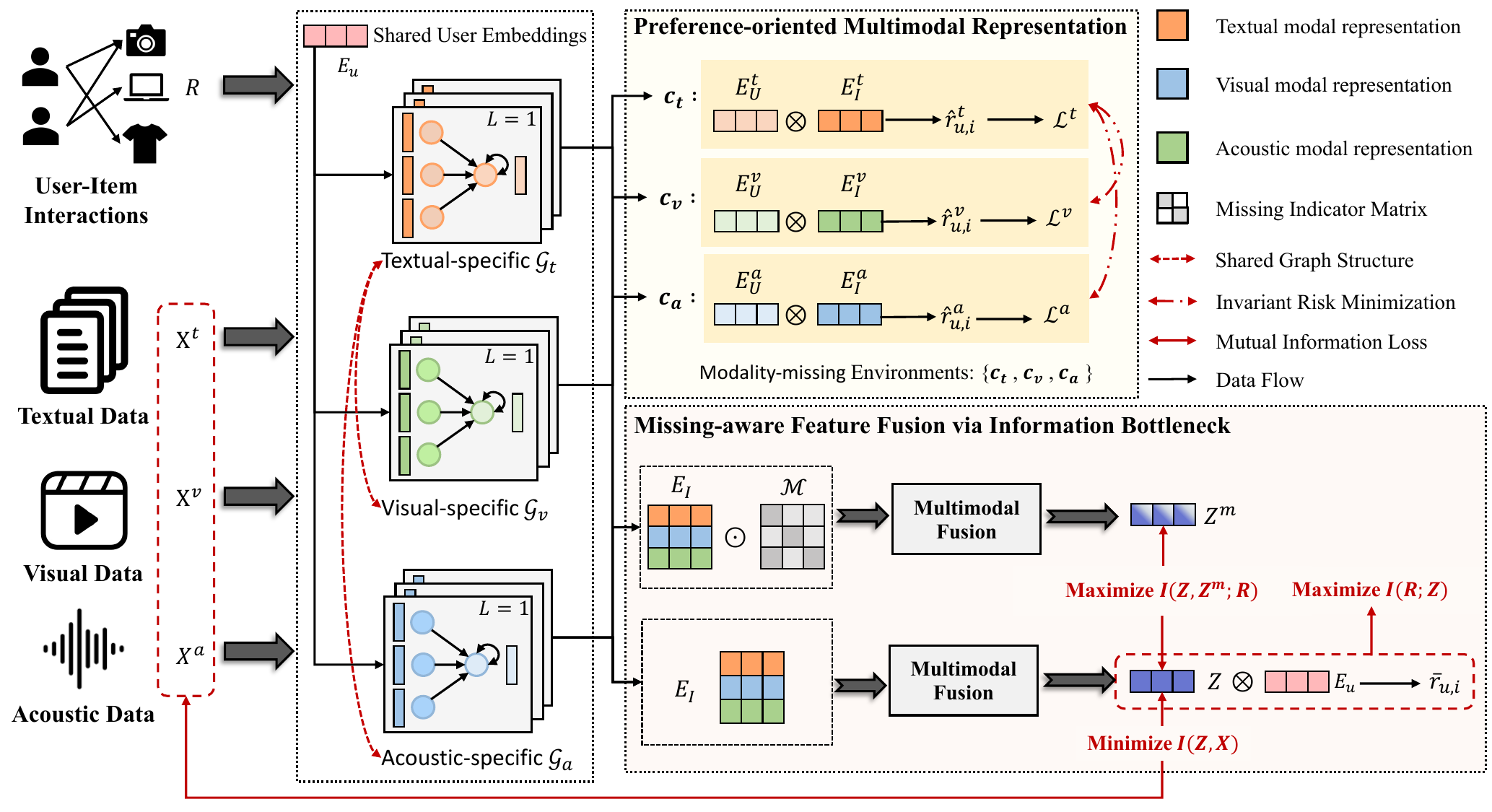}
    \vspace{0.0cm}
    \caption{Overview of the proposed I$^3$-MRec framework. The model first learns preference-oriented user/item representations using a graph-based approach, guided by IRM to ensure invariance across modalities. Then, an IB-based fusion module generates compact yet effective representations by maximizing preference-relevant information while compressing modality-specific information.}
    \label{fig:framework}
    \vspace{-0.2 cm}
\end{figure*}

\subsection{Multimodal Recommendation Task}
Let $\mathcal{U}$ and $\mathcal{I}$ denote the sets of users and items, with $|\mathcal{U}|= N_u $ and $|\mathcal{I}|= N_i $ representing the number of users and items. 
Items associated with multimodal features $\bm{X} \in \mathbb{R}^{N \times D_k}$, where $N$ is the number of modalities, and $k  \in \{v, t, a \}$ indexes modality types (visual, textual, and acoustic); $D_k$ is the dimension of $k$-th raw modality feature. The historical user-item interaction is denoted by the interaction matrix $\mathcal{R} \in \mathbb{R}^{|U| \times |I|} $, where  $r_{u,i} = 1 $ indicates an observed interaction between user $u$ and item $i$; otherwise, $r_{u,i} = 0 $. Thus, the multimodal recommendation task can be formally defined as predicting the probability $\bar{r}_{u, i}$ of user $u$ and item $i$:
\begin{equation}
\bar{r}_{u, i} = \Gamma(u,i, \bm{X}_i|\Theta).
\end{equation} 
Here, $\Gamma(\cdot)$ is any recommender learning function parameterized by $\Theta$ to predict preference probability. $\bm{X}_i$ represents the multimodal feature set of item $i$.

\subsection{Modality Missingness Problem}

MRS typically models user preferences by first extracting modality features and then fusing them into a unified item representation. Formally, this process can be described as:
\begin{equation}
\label{overall_function}
\bm{E}^k =g_{\theta_1}(\bm{X}^k), \qquad \bm{Z} = h_{\theta_2}(\bm{E}^a,\bm{E}^t,\bm{E}^v),
\end{equation}
where $g_{\theta_1}(\cdot)$ learns task-relevant semantic representations $\bm{E}^k$ from each modality input $\bm{X}^k$, and $h_{\theta_2}(\cdot)$ integrates them into a unified item representation $\bm{Z}$ for user preference prediction. The overall objective is defined as:
\begin{equation}
\mathcal{L}_0 = \mathcal{L}(p(\bm{E}_u, \bm{Z}), R),
\end{equation}
where $\bm{E}_u$ denotes user embeddings derived from user IDs\footnote{We adopt ID embeddings as user representations, assuming that user preferences are consistent across modalities.}, and $R$ is the ground-truth interaction label.

However, this modeling paradigm heavily depends on the availability of complete multimodal inputs. When certain modalities are missing, both the feature extraction and fusion stages become unreliable, resulting in degraded preference modeling. To mitigate this, we propose a robust framework that integrates invariant learning and the information bottleneck principle.

\section{Methodology}

In this section, we present I$^3$-MRec (Invariant learning with Information Bottleneck for Incomplete-Modality Recommendation), as illustrated in Fig.~\ref{fig:framework}, which consists of \textbf{Preference-oriented Multimodal Representation Learning } and \textbf{Missing-aware Feature Fusion via Information Bottleneck }. Next, we provide a detailed description of the overall optimization procedure.

\subsection{Preference-oriented Multimodal Representation Learning}
Existing invariant learning methods~\cite{ahuja2020invariant, du2022invariant, bai2024multimodality} commonly construct environments via synthetic distribution shifts, such as spurious correlations or artificial noise. In contrast, modalities in multimodal recommendation (e.g., text, image) naturally encode diverse yet complementary semantics. Given the consistency of user preferences across modalities~\cite{du2022invariant}, we treat each modality as a distinct environment to exploit this semantic invariance. 

\subsubsection{Collaborative Modality Graph Learning}
Formally, given $K$ modalities, we define the environment set $\mathcal{C} = \{c_k:1\leq k \leq K\}$, where $c_k$ represents the scenario where only the $k$-th modality is available. Then, we adopt modality-specific graphs to perform the representation learning of users and items. From the interaction matrix $\mathcal{R}$, we construct user-item interaction graph $\mathcal{G}=\{\mathcal{G}_a,\mathcal{G}_t,\mathcal{G}_v\}$. Each graph $\mathcal{G}_k$ maintains the same graph structure and only retains the item features associated with each modality. Note that all modality-specific graphs share the same user embeddings for each user ID~\cite{wei2019mmgcn}.

Given the raw modality features, we first employ a non-linear transformation to project each raw modality feature into a low-dimensional vector space:
\begin{equation}
\bm{F}^k = \sigma(\bm{W}_k \bm{X}^k+b_k),
\end{equation} 
where $\bm{W}_k \in \mathbb{R}^{D_k \times d_k } $ and $b_k \in \mathbb{R}^{d_k}$ denote the weight matrices and bias vectors for the $k-$th modality, respectively.  $\sigma(\cdot)$ is the activation function. To capture high-order collaborative signals, we employ the same message-passing strategy as LightGCN~\cite{He@LightGCN} to propagate and refine user and item representations across the user-item interaction graph. The $l$-th layer propagation is defined as:
\begin{equation}
\bm{E}^{(k, l)} =(\bm{D}^{-1/2}\bm{A}\bm{D}^{-1/2})\bm{E}^{(k, l-1)},
\end{equation}
where $\bm{A}$ is the adjacency matrix built from user-item interactions $\mathcal{R}$ and its transpose $\mathcal{R}^T$, and $\bm{D}$ is the corresponding diagonal degree matrix. By aggregating node features across $L$ propagation layers, the final preference-oriented embeddings can be obtained:
\begin{equation}
\bm{E}^ {k} = \frac{1}{L+1} \sum_{l=0}^L \bm{E}^{(k, l)}.
\end{equation}
Here, $\bm{E}^{k}=\{\bm{E}^k_U, \bm{E}^k_I\}$ represents preference-oriented representations of users and items. Based on the learned representations, we can estimate the preference score of the user $u$ for item $i$ in the $k$-th modality environment:
\begin{equation}
\hat{r}_{u, i}^k = \mathcal{F}(e_u^k, e_i^k),
\end{equation}
where $\mathcal{F}(\cdot)$ denotes the prediction function, implemented as the inner product between user and item embeddings. 

\subsubsection{Invariant Learning Optimization}
After obtaining predictions across $K$ modality environments, we explicitly encourage the consistency and stability of the learned representations across modalities. 
Specifically, we employ Invariant Risk Minimization ~\cite{arjovsky2019invariant, ahuja2020invariant, du2022invariant} to guide the learning of preference-oriented representations that remain invariant across environments. The corresponding optimization objective is formulated as follows:
\begin{equation}
\mathcal{L}_{IRM} =\mathbb{E}_{k\in \{a,t,v\}}\mathcal{L}^k+ \delta \lVert Var_{k\in \{a,t,v\}}(\nabla_{\theta_1}\mathcal{L}^k) \rVert^2.   
\end{equation} 
The first term, $\mathcal{L}^k$, represents the modality-specific loss for each environment. The second term penalizes the variance of the gradients (with respect to the shared parameters $\theta_1$) between different modalities. Minimizing this gradient variance encourages the model to capture invariant preference features that are stable across modality environments, thereby reducing its sensitivity to missing modalities.

To ensure effective learning of user preferences, we adopt the Bayesian Personalized Ranking (BPR) loss as the modality-specific objective:
\begin{equation}
\mathcal{L}^k = \sum_{(u,i^+, i^-) \in \mathcal{O} } -ln\sigma(\hat{r}_{u, i^+}^k - \hat{r}_{u, i^-}^k),
\end{equation} 
where $\mathcal{O} = {(u,i^+,i^-)|(u,i^+) \in \mathcal{R}^+, (u,i^-) \in \mathcal{R}^-}$ is the training triplet set, $(u,i^+)$ represents observed (positive) interactions, and $(u,i^-)$ denotes randomly sampled negative interactions. $\sigma(\cdot)$ is the sigmoid activation function. Through this design, our method ensures stable and accurate user preference prediction even in modality-missing scenarios.

\subsection{Missing-aware Feature Fusion}
To ensure robust performance under modality-missing scenarios, we design a missing-aware feature fusion module grounded in the Information Bottleneck (IB) principle, as illustrated in the lower part of Figure~\ref{sec:method}. As discussed above, modality-specific information can become unreliable in such scenarios, which may hinder accurate user preference prediction. To address this, the module is guided to preserve preference-relevant information while compressing modality-specific information, thereby generating compact yet effective multimodal representations.

\subsubsection{\textbf{Modality-Missing Scenarios Simulation}}
Instead of directly integrating all modality features, we introduce a binary mask to simulate modality-missing scenarios, which is dynamically sampled during training. Specifically, the mask vector $\mathcal{M} \in \{ 0,1 \}^K$ is applied to replace the features of missing modalities with zero vectors, where $m_k = 0$ indicates that the $k$-th modality is absent. The resulting masked representation is then fed into a single-layer MLP-based fusion function to generate the final item embedding:
\begin{equation}
z_{i, m} = h_{\theta_2}(\{e_i^a, e_i^t, e_i^v\}_m), \quad m \in \mathcal{M}.
\end{equation}
This strategy exposes the proposed model to various modality-missing scenarios during training, thus encouraging the fusion module to generate robust multimodal representations that ensure stable and consistent recommendation performance across different modality-missing conditions.

\subsubsection{\textbf{Information Bottleneck-guided Feature Fusion}}
Given generated representations, we employ the Information Bottleneck principle~\cite{tishby2015deep, tishby2000information, hu2024survey} to guide the optimization of the feature fusion module. It explicitly balances mutual information maximization and minimization, encouraging the model to retain task-relevant signals while compressing modality-specific information. Inspired by previous works~\cite{wei2022contrastive,liu2023debiased, yang2024graph}, the Information Bottleneck optimization objective in recommender systems can be formulated as:

\begin{equation}
IB = I(Z;R) + I(Z,Z^m;R) - \beta I(X;Z), 
\end{equation} 
here, $I(Z;R)$ encourages the model to preserve information relevant to the recommendation task; $I(Z,Z^m;R)$ promotes consistency across representations generated under different modality-missing conditions; and $I(X;Z)$ penalizes excessive dependence on raw modality content, controlled by a trade-off parameter $\beta$.

\textbf{Term1: Maximizing Mutual Information with Recommendation Signals. }
This term aims to maximize the mutual information between generated representations and the recommendation target. Given the user embedding $e_u$ and the generated representation $z_i$, the predicted preference score is computed as $\bar{r}_{u, i}^k = e_u^T, z_i$. We select popular BPR ranking loss ~\cite{rendle2009bpr} to optimize this term:
\begin{equation}
\mathcal{L}_{rec} = \sum_{(u,i^+, i^-) \in \mathcal{O} } -ln\sigma(\bar{r}_{u, i^+} - \bar{r}_{u, i^-}),
\end{equation} 

\begin{table}[t]
	\centering
	\caption{ Basic statistics of the experimental datasets.}
	\label{tab:data}
\resizebox{0.45\textwidth}{!}{
        \begin{tabular}{cccccl}
            \hline
            Dataset  & \#User & \multicolumn{1}{c}{\#Item} & \#interactions & \#Modalities & Sparsity \\ \hline
            Baby     & 19,445 & 7,050                       & 160.792        & V,T          & 99.88\%  \\
            Clothing & 39,387 & \multicolumn{1}{c}{23,033} & 278,677        & V,T          & 99.97\%  \\
            Tiktok   & 9,319  & 6,710                       & 68,722         & V,A,T        & 99.89\%  \\ \hline
        \end{tabular}}
	\vspace{-0.2cm}
\end{table}

\textbf{Term2: Promoting Robust Item Representation Consistency.} 
In general, enforcing consistency across different views has been shown to improve performance in downstream tasks~\cite{wan2021multi}. Motivated by this, we encourage semantic alignment between item representations generated under different masking conditions. Following standard contrastive learning setups~\cite{oord2018representation,sun2019infograph}, we employ the InfoNCE loss~\cite{gutmann2010noise} to estimate the mutual information $I(Z, Z^m; R)$:
\begin{equation}
\mathcal{L}_{pcon} = \sum_{\substack{{m,m'} \in \mathcal{M},\\ m \ne m'}} \sum_{i \in \mathcal{I}} -\log \frac{exp( \mathcal{s}(z_{i,m} \cdot z_{i,m'})/\tau)}{\sum_{j \in \mathcal{B}} exp(\mathcal{s}(z_{i,m} \cdot z_{j,m})/ \tau) },
\end{equation} 
where $\mathcal{s}(\cdot)$ measures the similarity between two representations, we employ the widely used cosine similarity function in our work; $\tau \in \mathbb{R}^+ $ indicates the temperature parameter. Note that negative examples of each item are updated at each epoch due to item $j$ being sampled from the current batch training data.

\textbf{Term3: Minimizing Mutual Information for Redundancy Reduction}. 
This term aims to compress the negative impact of modality-specific information by minimizing the mutual information between the generated representation and the raw modality input. In other words, it reduces the model’s reliance on potentially unreliable modality information. However, directly computing mutual information between two high-dimensional variables is intractable. To address this, we adopt the Contrastive Log-ratio Upper Bound (CLUB)~\cite{cheng2020club} in our work, which uses a variational distribution $q_{\phi}^k(\cdot | \cdot)$ for each modality to approximate the mutual information:
\begin{equation}
\mathcal{L}_{comp} = \frac{1}{|K|} \sum_{k \in \{a,t,v\}}{\sum_{i \in \mathcal{I} } \left[ \log q_{\phi^k}^k(z_i | e_i^k) - \log q_{\phi^k}^k(z_i | x_i^k)  \right] }
\end{equation} 
here, $q_{\phi}^k(\cdot | \cdot) $ is implemented as a two-layer MLP with modality-specific parameter $\phi^k$, optimized in a sample-based manner. Minimizing $\mathcal{L}^{comp}$ effectively reduces the task-irrelevant and modality information, yielding a more compact and effective representation. 

\textbf{Model Training}
The overall training objective optimization of our framework is defined as:
\begin{equation}
\mathcal{L} = \mathcal{L}_{rec} + \mathcal{L}_{IRM} + (\alpha\mathcal{L}_{pcon} + \beta\mathcal{L}_{comp}).   
\end{equation} 
This unified loss function integrates four components: the recommendation loss ($\mathcal{L}_{rec}$), the invariant representation learning loss ($\mathcal{L}_{IRM}$), the preference consistency loss ($\mathcal{L}_{pcon}$), and the information compression loss ($\mathcal{L}_{comp}$). The hyperparameters $\alpha$ and $\beta$ control the relative importance of the mutual information regularization terms. By jointly optimizing these objectives, the model learns compact and effective preference representations that generalize well to modality-missing scenarios.
\section{Experiments}
\label{sec:experiments}

\subsection{Experimental Setup}
\subsubsection{\textbf{Datasets}} 
We evaluate our method on three widely used real-world datasets: Amazon Baby, Amazon Clothing, and TikTok, following the 5-core setting adopted in prior works~\cite{BM32020Arxiv, Wei2020GRCN, ong2025spectrum}. The Amazon datasets\footnote{http://jmcauley.ucsd.edu/data/amazon.} consist of user-item interactions derived from product ratings. For each item, we extract 4096-dimensional visual features using VGG16~\cite{Simonyan2015VeryDC} (from the second fully connected layer), and 1024-dimensional textual features using BERT~\cite{Devlin2019BERTPO}, based on the concatenation of brand, title, description, and category, following~\cite{liu2023semantic}.
The TikTok dataset~\cite{bai2024multimodality} contains user viewing histories on short videos, with pre-extracted visual, acoustic, and textual features. We directly use the released modality features and data splits to ensure consistency and fair comparison. Following~\cite{liu2023semantic}, we adopt the same training, validation, and test partition strategy. Dataset statistics are summarized in Table~\ref{tab:data}.


\begin{table*}[t]
	\vspace{0pt}
	\caption{Performance of our method and the competitors over three datasets.} 
	\centering
    \vspace{-0.2cm}
 \resizebox{1.0\textwidth}{!}{
\begin{tabular}{l|cccccc|cccccc}
\hline
Settings & \multicolumn{6}{c|}{Full Training Missing Test (FTMT)}                                                                                              & \multicolumn{6}{c}{Missing Training Missing Test (MTMT)}                                                                                            \\ \hline
Datasets & \multicolumn{2}{c|}{Baby}                              & \multicolumn{2}{c|}{Clothing}                          & \multicolumn{2}{c|}{Tiktok}       & \multicolumn{2}{c|}{Baby}                              & \multicolumn{2}{c|}{Clothing}                          & \multicolumn{2}{c}{Tiktok}        \\ \hline
Metric   & Recall@20       & \multicolumn{1}{c|}{NDCG@20}         & Recall@20       & \multicolumn{1}{c|}{NDCG@20}         & Recall@20       & NDCG@20         & Recall@20       & \multicolumn{1}{c|}{NDCG@20}         & Recall@20       & \multicolumn{1}{c|}{NDCG@20}         & Recall@20       & NDCG@20         \\ \hline
MFBPR    & 0.0554          & \multicolumn{1}{c|}{0.0237}          & 0.0346          & \multicolumn{1}{c|}{0.1256}          & 0.0558          & 0.0220          & 0.0554          & \multicolumn{1}{c|}{0.0237}          & 0.0346          & \multicolumn{1}{c|}{0.1256}          & 0.0558          & 0.0220          \\
InvPref & 0.0562          & \multicolumn{1}{c|}{0.0241}          & 0.0356         & \multicolumn{1}{c|}{0.0147}          & 0.0572          & 0.0272         & 0.0562           & \multicolumn{1}{c|}{0.0241}          & 0.0356          & \multicolumn{1}{c|}{0.0147}          & 0.0572          & 0.0272          \\
LightGCN & 0.0687          & \multicolumn{1}{c|}{0.0320}          & 0.0436          & \multicolumn{1}{c|}{0.0145}          & 0.0736          & 0.0382          & 0.0687          & \multicolumn{1}{c|}{0.0320}          & 0.0436          & \multicolumn{1}{c|}{0.0149}          & 0.0736          & 0.0382          \\ \hline
VBPR     & 0.0472          & \multicolumn{1}{c|}{0.0205}          & 0.0462          & \multicolumn{1}{c|}{0.0207}          & 0.0406          & 0.0170          & 0.0433         & \multicolumn{1}{c|}{0.0162}          & 0.0413          & \multicolumn{1}{c|}{0.0184}          & 0.0359          & 0.0150          \\
MMGCN    & 0.0527          & \multicolumn{1}{c|}{0.0218}          & 0.0289          & \multicolumn{1}{c|}{0.012}           & 0.0874          & 0.0368          & 0.0402          & \multicolumn{1}{c|}{0.0165}          & 0.0258          & \multicolumn{1}{c|}{0.0107}          & 0.0753          & 0.0306          \\
GRCN     & 0.0602          & \multicolumn{1}{c|}{0.0255}          & 0.0381          & \multicolumn{1}{c|}{0.0161}          & 0.0709          & 0.0280          & 0.0572          & \multicolumn{1}{c|}{0.0242}          & 0.0340          & \multicolumn{1}{c|}{0.0144}          & 0.0602          & 0.0236          \\
DualGNN  & 0.0622          & \multicolumn{1}{c|}{0.0275}          & 0.0432          & \multicolumn{1}{c|}{0.0193}          & 0.0716          & 0.0340          & 0.0538          & \multicolumn{1}{c|}{0.0247}          & 0.0386          & \multicolumn{1}{c|}{0.0172}          & 0.0603          & 0.0297          \\

LATTICE  & 0.0632          & \multicolumn{1}{c|}{0.0278}          & 0.0511          & \multicolumn{1}{c|}{0.0215}          & 0.0743          & 0.0324          & 0.0542          & \multicolumn{1}{c|}{0.0241}          & 0.0437          & \multicolumn{1}{c|}{0.0187}          & 0.0627          & 0.0284          \\

InRL  & 0.0663          & \multicolumn{1}{c|}{0.0284}          & 0.0568          & \multicolumn{1}{c|}{0.0237}          & 0.0752          & 0.0321          & 0.0551          & \multicolumn{1}{c|}{0.0237}          & 0.0421          & \multicolumn{1}{c|}{0.0178}          & 0.0618          & 0.0276          \\

BM3      & 0.0683          & \multicolumn{1}{c|}{0.0296}          & 0.0572          & \multicolumn{1}{c|}{0.0252}          & 0.0760          & 0.0319          & 0.0611          & \multicolumn{1}{c|}{0.0257}          & 0.0510          & \multicolumn{1}{c|}{0.0225}          & 0.0670          & 0.0287          \\
MGCN     & 0.0788          & \multicolumn{1}{c|}{0.0322}          & 0.0591          & \multicolumn{1}{c|}{0.0268}          & 0.0861          & 0.0352          & 0.0697          & \multicolumn{1}{c|}{0.0272}          & 0.0527          & \multicolumn{1}{c|}{0.0239}          & 0.0745          & 0.0310          \\
SMORE    & 0.0792          & \multicolumn{1}{c|}{0.0332}          & \underline{0.0610}          & \multicolumn{1}{c|}{\underline{0.0278}  }          & \underline{0.0903}           & 0.0372          & 0.0712          & \multicolumn{1}{c|}{0.0308}          & \underline{0.0540 }         & \multicolumn{1}{c|}{ \underline{0.0247}}          & 0.0756          & 0.0316          \\ \hline
MILK     & 0.0415          & \multicolumn{1}{c|}{0.0184}          & 0.0226          & \multicolumn{1}{c|}{0.009}           & 0.0399          & 0.0182          & 0.0375          & \multicolumn{1}{c|}{0.0168}          & 0.0202          & \multicolumn{1}{c|}{0.0084}          & 0.0352          & 0.0157          \\
SIBRAR   & 0.0472          & \multicolumn{1}{c|}{0.0217}          & 0.0264          & \multicolumn{1}{c|}{0.011}           & 0.0542          & 0.0218          & 0.0429          & \multicolumn{1}{c|}{0.0198}          & 0.0236          & \multicolumn{1}{c|}{0.0098}          & 0.0468          & 0.0190          \\
MoDiCF   & \underline{0.0798}           & \multicolumn{1}{c|}{\underline{0.0348}}          & 0.0602          & \multicolumn{1}{c|}{0.0273}          & 0.0902          & \underline{0.0385}             & \underline{0.0724 }          & \multicolumn{1}{c|}{\underline{0.0318}}          & 0.0537          & \multicolumn{1}{c|}{0.0244}          & \underline{0.0775}         & \underline{ 0.0329}          \\ \hline
Ours     & \textbf{0.0815*} & \multicolumn{1}{c|}{\textbf{0.0361*}} & \textbf{0.0635*} & \multicolumn{1}{c|}{\textbf{0.0285*}} & \textbf{0.0929*} & \textbf{0.0398*} & \textbf{0.0744*} & \multicolumn{1}{c|}{\textbf{0.0325*}} & \textbf{0.0551*} & \multicolumn{1}{c|}{\textbf{0.0254*}} & \textbf{0.0804*} & \textbf{0.0343*} \\

Improv.  & 2.10\%          & \multicolumn{1}{c|}{3.62\%}          & 3.94\%          & \multicolumn{1}{c|}{2.49\%}          & 2.76\%          & 3.65\%          & 2.70\%          & \multicolumn{1}{c|}{2.17\%}          & 2.14\%          & \multicolumn{1}{c|}{3.54\%}          & 3.63\%          & 4.22\%          \\ \hline
\end{tabular}
 }
	\begin{tablenotes}
		\footnotesize
		\item The symbol * denotes that the improvement is significant with $p-value < 0.05$ based on a two-tailed paired t-test.
	\end{tablenotes}
	\label{tab:results}
	\vspace{0pt}
\end{table*}
\subsubsection{\textbf{Missing Modality Setting}} Similar to previous works ~\cite{lin2023contrastive, bai2024multimodality,li2025generating}, we evaluate the effectiveness of our method under missing modality scenarios using the following experimental settings. 
\begin{itemize}[leftmargin=0.35cm,topsep=3pt]
\item \textbf{Full Training Missing Test (FTMT)}: In this setting, all training data are assumed to have complete modality information. During testing, we simulate missing modalities by randomly selecting 50\% of the test items and replacing their modality features with zero vectors. For datasets with two modalities, each selected item may randomly lose one or both modalities. For datasets with three modalities, each item is randomly assigned one or two missing modalities to simulate diverse missing scenarios. 
\item \textbf{Missing Training Missing Test (MTMT)}: This setting simulates a more realistic scenario in which missing modalities also occur during training. Specifically, 30\% of the training items are randomly assigned missing modalities, using the same missing strategy as in the test phase. The test phase follows the same setup as FTMT.
\end{itemize}

\subsubsection{\textbf{Baselines}}
To verify the efficacy of our proposed method, we conduct comprehensive benchmarking against a diverse set of state-of-the-art (SOTA) recommender models, which can be broadly categorized into three main groups: \textbf{Unimodal Recommendation Methods}: MF-BPR ~\cite{rendle2009bpr}, InvPref~\cite{wang2022invariant}, and LightGCN~\cite{He@LightGCN}. These methods rely solely on user-item interaction data to infer user preferences, without incorporating auxiliary content features. \textbf{Multimodal Recommendation Methods}: feature-based models (VBPR~\cite{he2016vbpr}, BM3~\cite{BM32020Arxiv}, InRL~\cite{du2022invariant}), graph-based models (GRCN~\cite{Wei2020GRCN}, MMGCN~\cite{wei2019mmgcn},  LATTICE~\cite{zhang2021mining}, DualGNN~\cite{wang2021dualgnn}), and hybrid models (MGCN~\cite{yu2023multi}, SMORE~\cite{ong2025spectrum}). \textbf{Incomplete Modality Recommendation Methods:} These models are specifically designed to alleviate modality-missing problems, which include representative approaches (MILK~\cite{bai2024multimodality}, SIBRAR~\cite{ganhor2024multimodal}), and Generative models(MoDiCF~\cite{li2025generating}).

\subsubsection{\textbf{Evaluation Metrics}}
Two widely-used evaluation metrics are adopted for top-$n$ recommendation: \emph{Recall} and \emph{Normalized Discounted Cumulative Gain} (NDCG)~\cite{he2015trirank}. 
Recommendation accuracy is calculated for each metric based on the top 20 results. Note that the reported results are the average values across all testing users. 

\subsubsection{\textbf{Parameter Settings}}
The PyTorch framework~\footnote{https://pytorch.org.} is adopted to implement the proposed model. In our experiments, all hyperparameters are carefully tuned. For all datasets, the embedding size of the users and items is set to 64. The mini-batch size is fixed to 1024. The learning rate for the optimizer is searched from $\{10^{-5},10^{-4},\cdots,10^{+1}\}$, and the model weight decay is searched in the range $\{10^{-5}, 10^{-4},\cdots, 10^{-1}\}$. In the mutual information $I(Z,Z^m;R)$, the temperature $\tau$ is tuned from$\{1e-2,1e-1,1,1e+1\}$. We carefully searched the best parameters of $\alpha$, $\beta$, and $\delta$, and found that I$^3$-MRec achieves the best performance when $\{1e-3, 1e-5, 300 \}$ on Baby, $\{1e-2, 1e-3, 1 \}$ on Clothing, and $\{1e-3, 1e-6, 1000 \}$ on TikTok dataset. Besides, the early stopping strategy is adopted. Specifically, the training process will stop if recall@20 does not increase for 20 successive epochs. 

\subsection{\textbf{Overall Comparisons}}

Table~\ref {tab:results} presents the performance comparison between I$^3$-MRec and several SOTA baselines across three datasets. To better simulate real-world scenarios where modality information may be randomly missing during both training and inference, we conduct evaluations under two experimental settings: Full Training Missing Test (FTMT) and Missing Training Missing Test (MTMT). The key observations are summarized as follows:

The first block reports the performance of unimodal recommender methods, including MFBPR, InvPref, and LightGCN, which are unaffected by modality missingness. InvPref achieves noticeable improvements over MFBPR, demonstrating the effectiveness of invariant learning in enhancing robustness and generalization. LightGCN further outperforms both MFBPR and InvPref by leveraging high-order neighborhood information to enrich user and item representations, underscoring the strengths of graph-based methods in mitigating data sparsity and improving representation quality. Building upon a similar GNN-based architecture, the proposed I$^3$-MRec additionally integrates invariant learning to learn robust multimodal representations. 

The second block presents the performance of multimodal recommender methods. As shown, most MRS methods outperform unimodal baselines by leveraging richer semantic inputs. However, since they are not specifically designed for modality-missing scenarios, their performance remains sensitive to modality missingness. For instance, MMGCN relies on complete modality for node initialization and message passing. When modalities are missing during testing (FTMT), their performance drops below that of unimodal models. Introducing missing modalities during training (MTMT) results in a substantial performance drop of up to 23\%, indicating the model’s vulnerability to modality-missing problems across the learning process. In contrast, I$^3$-MRec explicitly accounts for modality missingness during training, yielding consistently strong performance across various modality-missing scenarios. InRL also adopts the invariant learning paradigm to learn preference-invariant representations. However, it relies on raw multimedia content to infer interaction environments, which becomes unreliable under missing modalities, causing performance degradation of up to 25\%. These findings highlight the importance of explicitly alleviating modality-missing problems in multimodal recommendations.

In the third block, we analyze the performance of missing modality-aware recommender methods. Both SiBraR and MILK are primarily designed for cold-start scenarios, and thus do not fully exploit collaborative filtering signals~\cite{ganhor2024multimodal}. This partly explains their lower performance compared to unimodal baselines. Nevertheless, these models show strong robustness to missing modalities. For example, on the Baby dataset (top-20), SiBraR's performance only drops from 0.0472 (FTMT) to 0.0429 (MTMT), a decrease of just 9.1\%, indicating better stability than other multimodal baselines in modality-missing scenarios. Second, MoDiCF performs well on the Baby dataset due to its effective generation and debiasing design, but its performance drops notably on TikTok and Clothing. This is because TikTok’s uneven three-modality structure and weak cross-modal correlation hinder reliable imputation, while Clothing depends heavily on visual features, making image absence particularly damaging. In contrast, I$^3$-MRec avoids reconstruction and achieves consistently strong results across all datasets and settings (FTMT and MTMT).

These observations highlight the inherent limitations of existing multimodal methods in realistic modality-missing scenarios. In contrast, our proposed method consistently delivers stable and superior performance across all datasets and settings, demonstrating its effectiveness and robustness.

\begin{figure}[t]
    \vspace{-0pt}
    \centering
    \subfloat[]{\includegraphics[width=0.5 \linewidth]{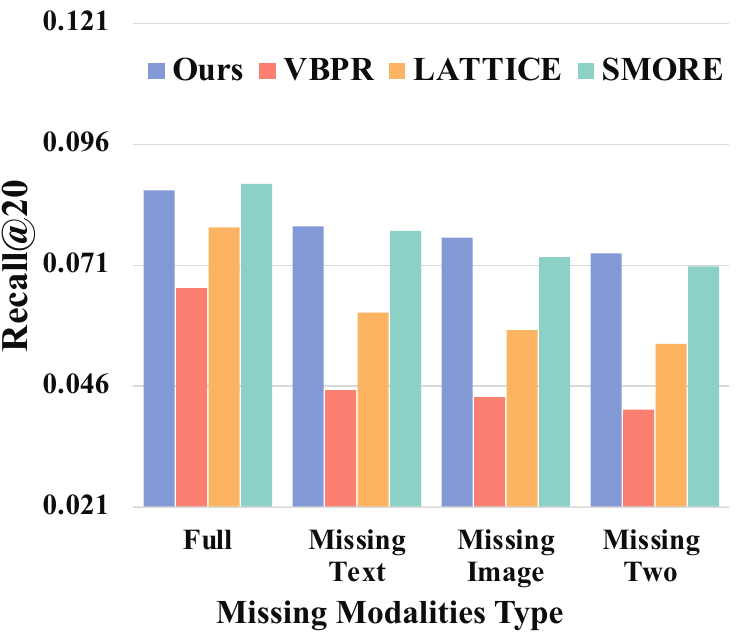}}
    \subfloat[]{\includegraphics[width=0.5 \linewidth]{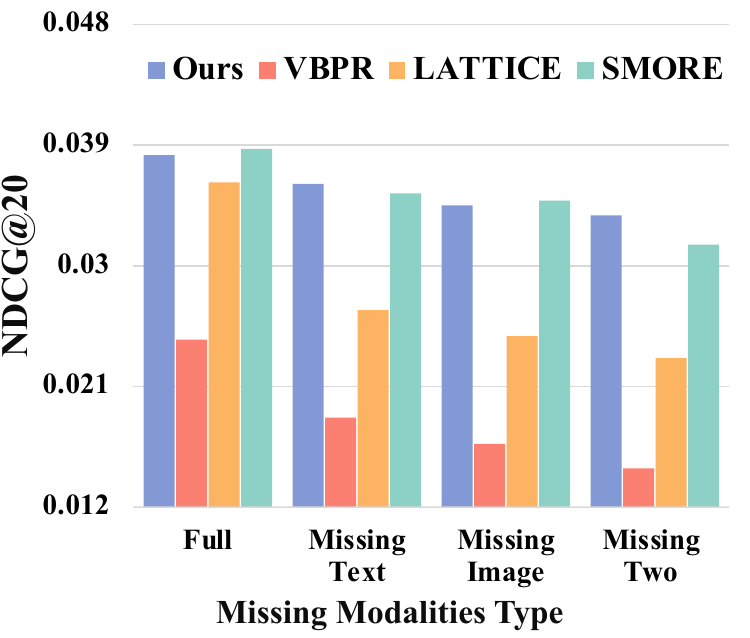}}
    \caption{Performance on different types of missing modality on the Amazon Baby dataset. The “Full” setting indicates that the baseline models were trained with complete modality scenarios. }
    \label{fig:missing_modalities_type}
    \vspace{- 0.2 cm}
\end{figure}

\subsection{\textbf{Improvements on Modality Missingness}}
\subsubsection{Performance at Different Missing Modality Types.}

In Figure ~\ref{fig:missing_modalities_type}, we evaluate the performance of the proposed model and MRS (e.g., VBPR, LATTICE, SMORE) across different types of missing modalities. To simulate the real-world scenario, we select the MTMT strategy as the experimental setting\footnote{The MTMT setting simulates realistic and diverse missing cases, including mismatches between training and test missing modalities (e.g., training with missing image, testing with missing text or both}.
From the observation of the figure, we can find that 1) I$^3$-MRec consistently outperformed all baselines across all modality-missing scenarios. This demonstrates the effectiveness of our framework in learning robust representations even in the absence of two modalities. 
2) Benefiting from the introduction of invariant learning and information bottleneck, I$^3$-MRec can yield compact yet effective representation, which maximally preserves user preference information while minimizing dependence on the raw modality content. As shown in the figure, our model maintains stable performance regardless of which modality is missing. while traditional MRS models are especially sensitive to the absence of visual features, leading to a sharp performance drop. 3) Beyond robustness under modality-missing conditions, I$^3$-MRec also performs competitively in full-modality settings, achieving results on par with the recent work SMORE. Interestingly, the motivation of SMORE is to capture both uni-modal and fusion preferences while simultaneously suppressing modality noise. This suggests that reducing modality dependence through the Information Bottleneck principle benefits generalization in modality-missing scenarios.

\begin{figure}[t]
    \vspace{0 pt}
    \centering

    \subfloat[]{\includegraphics[width=0.5 \linewidth]{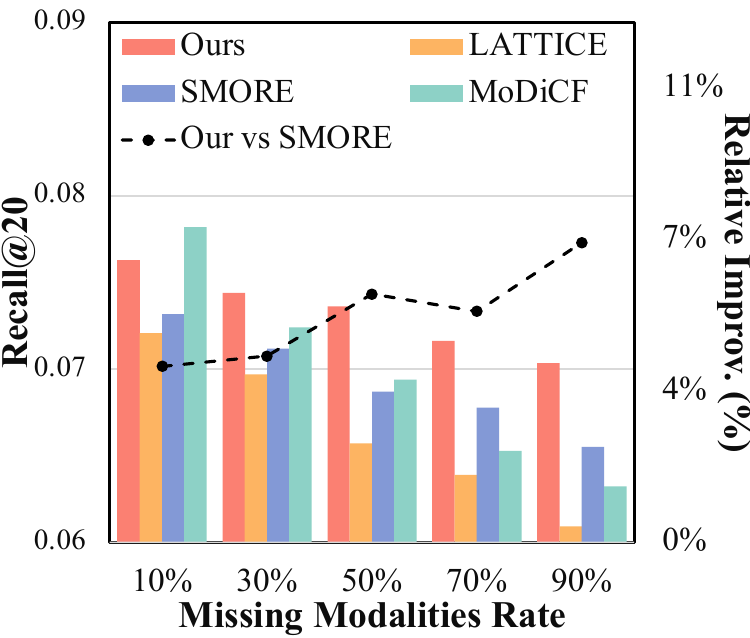}}
    \subfloat[]{\includegraphics[width=0.5 \linewidth]{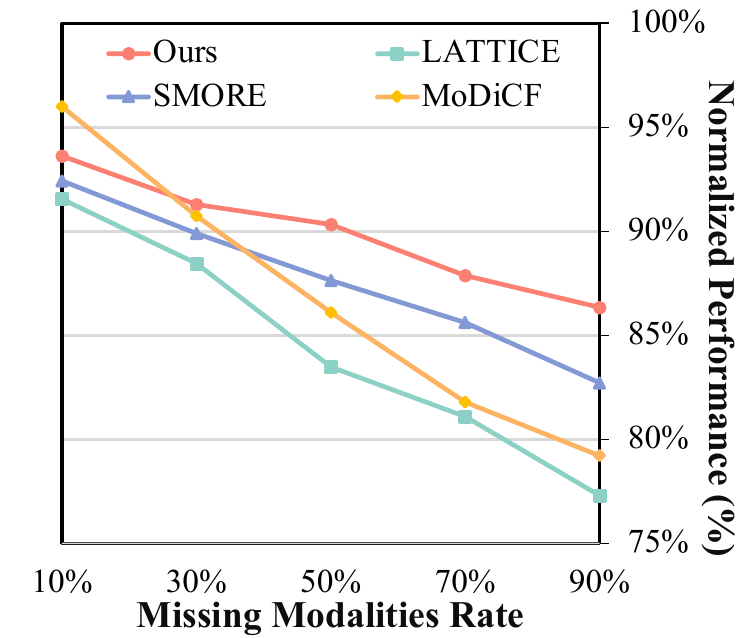}}
    
    \caption{(a) Recall@20 under varying missing modality rates on the Amazon Baby dataset. (b) Normalized performance with respect to the full-modality setting.
    }
    \label{fig:missing_modalities_rate}
    \vspace{-0.1 cm}
\end{figure}

\subsubsection{Performance under Varying Missing Ratios.}
To evaluate the robustness of I$^3$-MRec under varying degrees of modality missingness, we conduct experiments with missing ratios of $\{10\%, 30\%, 50\%,$ $70\%, 90\%\} $ in the MTMT setting. For each item, both the items with missing modalities and the types of missing modalities are randomly selected and fixed across all settings. For example, if an item’s textual modality is missing at a 10\% ratio, the same modality for that item remains missing at higher ratios.

Figure~\ref{fig:missing_modalities_rate}(a) reports the Recall@20 results of I$^3$-MRec compared with LATTICE, SMORE, and MoDiCF on the Amazon Baby dataset. I$^3$-MRec consistently outperforms all baselines across all missing ratios. While existing methods suffer substantial performance degradation as the missing rate increases, I$^3$-MRec remains notably stable. The relative improvement over the strongest baseline, SMORE, increases steadily with the missing ratio and reaches over 7\% at 90\%, highlighting the effectiveness of our method under both moderate and extreme modality-missing conditions. Figure~\ref{fig:missing_modalities_rate}(b) further presents the normalized performance with respect to the full-modality setting. I$^3$-MRec demonstrates the smallest performance drop, with only a 12\% degradation at a 90\% missing rate, clearly indicating superior robustness compared to all other methods.

Interestingly, MoDiCF performs well under low missing ratios, retaining 98\% of full-modality performance at 10\%. However, as missingness increases (50\%-90\%), its effectiveness declines sharply, with recall dropping by nearly 20\%, revealing the limitations of generation-based strategies in modality-missing scenarios. In contrast, I$^3$-MRec learns compact and effective representations, maintaining stable performance even when most modalities are unavailable, demonstrating the model’s robustness and generalization.

\subsection{\textbf{Ablation Study}}

Table~\ref{tab:ablation_studay} presents the ablation results of I$^3$-MRec under the FTMT and MTMT settings. We first analyze the individual contributions of the IRM and IB modules to user preference modeling in the presence of missing modalities. The variant \textbf{I$^3$-MRec$_\text{IRM}$}, which applies only invariant learning across modality-specific environments, outperforms the IB-only variant \textbf{I$^3$-MRec$_\text{IB}$}, demonstrating that environment-based invariant learning effectively enhances robustness under modality-missing conditions. IRM encourages consistency in user preference modeling across different modalities, leading to better generalization. 

While \textbf{I$^3$-MRec$_\text{IB}$} is less effective on its own, it still improves upon a naive baseline, confirming that the IB module contributes to robustness by guiding the model to learn compact yet effective representations focused on preference-relevant semantics, thereby reducing the impact of modality missingness.

To further examine the role of the IB module, we introduce \textbf{I$^3$-MRec$^*$}, a variant that incorporates both IRM and IB but omits the information compression (minimization) term. Its performance lags behind the full model, particularly under the MTMT setting, indicating that without information compression, the model becomes more susceptible to missing modality inputs during training.

\begin{table}[h]
\centering
\caption{ Ablation studies on the components of our proposed method.}
 \resizebox{0.45\textwidth}{!}{
\begin{tabular}{l|cc|cc}
\hline
Settings & \multicolumn{2}{c|}{FTMT} & \multicolumn{2}{c}{MTMT}                 \\ \hline \hline
Variants & Recall@20    & NDCG@20    & Recall@20 & \multicolumn{1}{c}{NDCG@20} \\ \hline
I$^3$-MRec$_{IB}$         & 0.0763      & 0.0338    & 0.0596   & \multicolumn{1}{c}{0.0254} \\
I$^3$-MRec$_{IRM}$          & 0.0799      & 0.0348    & 0.0638   & \multicolumn{1}{c}{0.0279}   \\
I$^3$-MRec$^*$        & 0.0809      & 0.0351    & 0.0727   & \multicolumn{1}{c}{0.0307}  \\
I$^3$-MRec                 & 0.0817      & 0.0361    & 0.0744   & \multicolumn{1}{c}{0.0325} \\ \hline \hline
\end{tabular}}
\label{tab:ablation_studay}
\vspace{0 pt}
\end{table}


\section{Conclusion}
\label{sec:conclusion}
In this work, we make a principled contribution to robust multimodal recommendation by explicitly identifying two fundamental properties essential for effective recommendation under missing modality conditions: (i) cross-modal preferences invariance and (ii) compact, preference-oriented representation. Grounded in these properties, we propose I$^3$-MRec, a novel and generalizable framework that integrates Invariant Risk Minimization (IRM) and the Information Bottleneck (IB) principle. Specifically, IRM enables the model to learn stable, reference-oriented representations across modality-specific environments, while IB encourages the model to preserve preference-relevant information and reduce reliance on modality-specific information. This design allows I$^3$-MRec to achieve robust and accurate recommendations in realistic modality-missing scenarios. Extensive experiments on three real-world datasets demonstrate that I$^3$-MRec consistently outperforms unimodal, multimodal, and modality-missing aware baselines across a range of missing ratios and evaluation settings (FTMT and MTMT). These results validate the robustness and generalization capabilities of our framework. 

\section{Acknowledgments}
This research is supported by the National Natural Science Foundation
of China (Nos. 62272254, 72188101, 62020106007, 62476071). And, the computation is completed on the HPC Platform of Hefei University of Technology. Any opinions, findings, and conclusions or recommendations expressed in this material are those of the author(s) and do not reflect the views of the National Natural Science Foundation of China.

\bibliographystyle{ACM-Reference-Format}
\bibliography{sample-sigconf}


\begin{thebibliography}{76}


\ifx \showCODEN    \undefined \def \showCODEN     #1{\unskip}     \fi
\ifx \showISBNx    \undefined \def \showISBNx     #1{\unskip}     \fi
\ifx \showISBNxiii \undefined \def \showISBNxiii  #1{\unskip}     \fi
\ifx \showISSN     \undefined \def \showISSN      #1{\unskip}     \fi
\ifx \showLCCN     \undefined \def \showLCCN      #1{\unskip}     \fi
\ifx \shownote     \undefined \def \shownote      #1{#1}          \fi
\ifx \showarticletitle \undefined \def \showarticletitle #1{#1}   \fi
\ifx \showURL      \undefined \def \showURL       {\relax}        \fi
\providecommand\bibfield[2]{#2}
\providecommand\bibinfo[2]{#2}
\providecommand\natexlab[1]{#1}
\providecommand\showeprint[2][]{arXiv:#2}

\bibitem[Ahuja et~al\mbox{.}(2020)]%
        {ahuja2020invariant}
\bibfield{author}{\bibinfo{person}{Kartik Ahuja}, \bibinfo{person}{Karthikeyan Shanmugam}, \bibinfo{person}{Kush Varshney}, {and} \bibinfo{person}{Amit Dhurandhar}.} \bibinfo{year}{2020}\natexlab{}.
\newblock \showarticletitle{Invariant risk minimization games}. In \bibinfo{booktitle}{\emph{International Conference on Machine Learning}}. \bibinfo{pages}{145--155}.
\newblock


\bibitem[Arjovsky et~al\mbox{.}(2019)]%
        {arjovsky2019invariant}
\bibfield{author}{\bibinfo{person}{Martin Arjovsky}, \bibinfo{person}{L{\'e}on Bottou}, \bibinfo{person}{Ishaan Gulrajani}, {and} \bibinfo{person}{David Lopez-Paz}.} \bibinfo{year}{2019}\natexlab{}.
\newblock \showarticletitle{Invariant risk minimization}.
\newblock \bibinfo{journal}{\emph{arXiv preprint arXiv:1907.02893}} (\bibinfo{year}{2019}).
\newblock


\bibitem[Bai et~al\mbox{.}(2024)]%
        {bai2024multimodality}
\bibfield{author}{\bibinfo{person}{Haoyue Bai}, \bibinfo{person}{Le Wu}, \bibinfo{person}{Min Hou}, \bibinfo{person}{Miaomiao Cai}, \bibinfo{person}{Zhuangzhuang He}, \bibinfo{person}{Yuyang Zhou}, \bibinfo{person}{Richang Hong}, {and} \bibinfo{person}{Meng Wang}.} \bibinfo{year}{2024}\natexlab{}.
\newblock \showarticletitle{Multimodality invariant learning for multimedia-based new item recommendation}. In \bibinfo{booktitle}{\emph{Proceedings of the 47th International ACM SIGIR Conference on Research and Development in Information Retrieval}}. \bibinfo{pages}{677--686}.
\newblock


\bibitem[Belghazi et~al\mbox{.}(2018)]%
        {belghazi2018mutual}
\bibfield{author}{\bibinfo{person}{Mohamed~Ishmael Belghazi}, \bibinfo{person}{Aristide Baratin}, \bibinfo{person}{Sai Rajeshwar}, \bibinfo{person}{Sherjil Ozair}, \bibinfo{person}{Yoshua Bengio}, \bibinfo{person}{Aaron Courville}, {and} \bibinfo{person}{Devon Hjelm}.} \bibinfo{year}{2018}\natexlab{}.
\newblock \showarticletitle{Mutual information neural estimation}. In \bibinfo{booktitle}{\emph{International conference on machine learning}}. \bibinfo{pages}{531--540}.
\newblock


\bibitem[Cai et~al\mbox{.}(2022)]%
        {cai2022adaptive}
\bibfield{author}{\bibinfo{person}{Desheng Cai}, \bibinfo{person}{Shengsheng Qian}, \bibinfo{person}{Quan Fang}, \bibinfo{person}{Jun Hu}, {and} \bibinfo{person}{Changsheng Xu}.} \bibinfo{year}{2022}\natexlab{}.
\newblock \showarticletitle{Adaptive anti-bottleneck multi-modal graph learning network for personalized micro-video recommendation}. In \bibinfo{booktitle}{\emph{Proceedings of the 30th ACM International Conference on Multimedia}}. \bibinfo{pages}{581--590}.
\newblock


\bibitem[Cai et~al\mbox{.}(2021)]%
        {cai2021heterogeneous}
\bibfield{author}{\bibinfo{person}{Desheng Cai}, \bibinfo{person}{Shengsheng Qian}, \bibinfo{person}{Quan Fang}, {and} \bibinfo{person}{Changsheng Xu}.} \bibinfo{year}{2021}\natexlab{}.
\newblock \showarticletitle{Heterogeneous hierarchical feature aggregation network for personalized micro-video recommendation}.
\newblock \bibinfo{journal}{\emph{IEEE Transactions on Multimedia}} (\bibinfo{year}{2021}), \bibinfo{pages}{805--818}.
\newblock


\bibitem[Chen et~al\mbox{.}(2017)]%
        {chen2017attentive}
\bibfield{author}{\bibinfo{person}{Jingyuan Chen}, \bibinfo{person}{Hanwang Zhang}, \bibinfo{person}{Xiangnan He}, \bibinfo{person}{Liqiang Nie}, \bibinfo{person}{Wei Liu}, {and} \bibinfo{person}{Tat-Seng Chua}.} \bibinfo{year}{2017}\natexlab{}.
\newblock \showarticletitle{Attentive collaborative filtering: Multimedia recommendation with item-and component-level attention}. In \bibinfo{booktitle}{\emph{Proceedings of the 40th International ACM SIGIR conference on Research and Development in Information Retrieval}}. \bibinfo{pages}{335--344}.
\newblock


\bibitem[Cheng et~al\mbox{.}(2020)]%
        {cheng2020club}
\bibfield{author}{\bibinfo{person}{Pengyu Cheng}, \bibinfo{person}{Weituo Hao}, \bibinfo{person}{Shuyang Dai}, \bibinfo{person}{Jiachang Liu}, \bibinfo{person}{Zhe Gan}, {and} \bibinfo{person}{Lawrence Carin}.} \bibinfo{year}{2020}\natexlab{}.
\newblock \showarticletitle{Club: A contrastive log-ratio upper bound of mutual information}. In \bibinfo{booktitle}{\emph{International conference on machine learning}}. \bibinfo{pages}{1779--1788}.
\newblock


\bibitem[Creager et~al\mbox{.}(2021)]%
        {creager2021environment}
\bibfield{author}{\bibinfo{person}{Elliot Creager}, \bibinfo{person}{J{\"o}rn-Henrik Jacobsen}, {and} \bibinfo{person}{Richard Zemel}.} \bibinfo{year}{2021}\natexlab{}.
\newblock \showarticletitle{Environment inference for invariant learning}. In \bibinfo{booktitle}{\emph{International Conference on Machine Learning}}. \bibinfo{pages}{2189--2200}.
\newblock


\bibitem[Ding et~al\mbox{.}(2023)]%
        {ding2023robust}
\bibfield{author}{\bibinfo{person}{Shifei Ding}, \bibinfo{person}{Wei Du}, \bibinfo{person}{Ling Ding}, \bibinfo{person}{Jian Zhang}, \bibinfo{person}{Lili Guo}, {and} \bibinfo{person}{Bo An}.} \bibinfo{year}{2023}\natexlab{}.
\newblock \showarticletitle{Robust multi-agent communication with graph information bottleneck optimization}.
\newblock \bibinfo{journal}{\emph{IEEE Transactions on Pattern Analysis and Machine Intelligence}} (\bibinfo{year}{2023}), \bibinfo{pages}{3096--3107}.
\newblock


\bibitem[Du et~al\mbox{.}(2022)]%
        {du2022invariant}
\bibfield{author}{\bibinfo{person}{Xiaoyu Du}, \bibinfo{person}{Zike Wu}, \bibinfo{person}{Fuli Feng}, \bibinfo{person}{Xiangnan He}, {and} \bibinfo{person}{Jinhui Tang}.} \bibinfo{year}{2022}\natexlab{}.
\newblock \showarticletitle{Invariant representation learning for multimedia recommendation}. In \bibinfo{booktitle}{\emph{Proceedings of the 30th ACM International Conference on Multimedia}}. \bibinfo{pages}{619--628}.
\newblock


\bibitem[Espinosa~Zarlenga et~al\mbox{.}(2022)]%
        {espinosa2022concept}
\bibfield{author}{\bibinfo{person}{Mateo Espinosa~Zarlenga}, \bibinfo{person}{Pietro Barbiero}, \bibinfo{person}{Gabriele Ciravegna}, \bibinfo{person}{Giuseppe Marra}, \bibinfo{person}{Francesco Giannini}, \bibinfo{person}{Michelangelo Diligenti}, \bibinfo{person}{Zohreh Shams}, \bibinfo{person}{Frederic Precioso}, \bibinfo{person}{Stefano Melacci}, \bibinfo{person}{Adrian Weller}, {et~al\mbox{.}}} \bibinfo{year}{2022}\natexlab{}.
\newblock \showarticletitle{Concept embedding models: Beyond the accuracy-explainability trade-off}.
\newblock \bibinfo{journal}{\emph{Advances in Neural Information Processing Systems}} (\bibinfo{year}{2022}), \bibinfo{pages}{21400--21413}.
\newblock


\bibitem[Ganh{\"o}r et~al\mbox{.}(2024)]%
        {ganhor2024multimodal}
\bibfield{author}{\bibinfo{person}{Christian Ganh{\"o}r}, \bibinfo{person}{Marta Moscati}, \bibinfo{person}{Anna Hausberger}, \bibinfo{person}{Shah Nawaz}, {and} \bibinfo{person}{Markus Schedl}.} \bibinfo{year}{2024}\natexlab{}.
\newblock \showarticletitle{A Multimodal Single-Branch Embedding Network for Recommendation in Cold-Start and Missing Modality Scenarios}. In \bibinfo{booktitle}{\emph{Proceedings of the 18th ACM Conference on Recommender Systems}}. \bibinfo{pages}{380--390}.
\newblock


\bibitem[Gronowski et~al\mbox{.}(2023)]%
        {gronowski2023classification}
\bibfield{author}{\bibinfo{person}{Adam Gronowski}, \bibinfo{person}{William Paul}, \bibinfo{person}{Fady Alajaji}, \bibinfo{person}{Bahman Gharesifard}, {and} \bibinfo{person}{Philippe Burlina}.} \bibinfo{year}{2023}\natexlab{}.
\newblock \showarticletitle{Classification utility, fairness, and compactness via tunable information bottleneck and R{\'e}nyi measures}.
\newblock \bibinfo{journal}{\emph{IEEE Transactions on Information Forensics and Security}} (\bibinfo{year}{2023}), \bibinfo{pages}{1630--1645}.
\newblock


\bibitem[Gutmann and Hyv{\"a}rinen(2010)]%
        {gutmann2010noise}
\bibfield{author}{\bibinfo{person}{Michael Gutmann} {and} \bibinfo{person}{Aapo Hyv{\"a}rinen}.} \bibinfo{year}{2010}\natexlab{}.
\newblock \showarticletitle{Noise-contrastive estimation: A new estimation principle for unnormalized statistical models}. In \bibinfo{booktitle}{\emph{Proceedings of the thirteenth international conference on artificial intelligence and statistics}}. \bibinfo{pages}{297--304}.
\newblock


\bibitem[He and McAuley(2016)]%
        {he2016vbpr}
\bibfield{author}{\bibinfo{person}{Ruining He} {and} \bibinfo{person}{Julian McAuley}.} \bibinfo{year}{2016}\natexlab{}.
\newblock \showarticletitle{VBPR: visual bayesian personalized ranking from implicit feedback}. In \bibinfo{booktitle}{\emph{Proceedings of the AAAI conference on artificial intelligence}}.
\newblock


\bibitem[He et~al\mbox{.}(2015)]%
        {he2015trirank}
\bibfield{author}{\bibinfo{person}{Xiangnan He}, \bibinfo{person}{Tao Chen}, \bibinfo{person}{Min-Yen Kan}, {and} \bibinfo{person}{Xiao Chen}.} \bibinfo{year}{2015}\natexlab{}.
\newblock \showarticletitle{Trirank: Review-aware explainable recommendation by modeling aspects}. In \bibinfo{booktitle}{\emph{Proceedings of the 24th ACM international on conference on information and knowledge management}}. \bibinfo{pages}{1661--1670}.
\newblock


\bibitem[He et~al\mbox{.}(2020)]%
        {He@LightGCN}
\bibfield{author}{\bibinfo{person}{Xiangnan He}, \bibinfo{person}{Kuan Deng}, \bibinfo{person}{Xiang Wang}, \bibinfo{person}{Yan Li}, \bibinfo{person}{Yongdong Zhang}, {and} \bibinfo{person}{Meng Wang}.} \bibinfo{year}{2020}\natexlab{}.
\newblock \showarticletitle{Lightgcn: Simplifying and powering graph convolution network for recommendation}. In \bibinfo{booktitle}{\emph{Proceedings of the 43rd International ACM SIGIR conference on research and development in Information Retrieval}}. \bibinfo{pages}{639--648}.
\newblock


\bibitem[Hu et~al\mbox{.}(2024)]%
        {hu2024survey}
\bibfield{author}{\bibinfo{person}{Shizhe Hu}, \bibinfo{person}{Zhengzheng Lou}, \bibinfo{person}{Xiaoqiang Yan}, {and} \bibinfo{person}{Yangdong Ye}.} \bibinfo{year}{2024}\natexlab{}.
\newblock \showarticletitle{A survey on information bottleneck}.
\newblock \bibinfo{journal}{\emph{IEEE Transactions on Pattern Analysis and Machine Intelligence}} (\bibinfo{year}{2024}).
\newblock


\bibitem[Kang et~al\mbox{.}(2017)]%
        {kang2017visually}
\bibfield{author}{\bibinfo{person}{Wang-Cheng Kang}, \bibinfo{person}{Chen Fang}, \bibinfo{person}{Zhaowen Wang}, {and} \bibinfo{person}{Julian McAuley}.} \bibinfo{year}{2017}\natexlab{}.
\newblock \showarticletitle{Visually-aware fashion recommendation and design with generative image models}. In \bibinfo{booktitle}{\emph{2017 IEEE International Conference on Data Mining}}. \bibinfo{pages}{207--216}.
\newblock


\bibitem[Kenton et~al\mbox{.}(2019)]%
        {Devlin2019BERTPO}
\bibfield{author}{\bibinfo{person}{Jacob Devlin Ming-Wei~Chang Kenton}, \bibinfo{person}{Lee~Kristina Toutanova}, {et~al\mbox{.}}} \bibinfo{year}{2019}\natexlab{}.
\newblock \showarticletitle{Bert: Pre-training of deep bidirectional transformers for language understanding}. In \bibinfo{booktitle}{\emph{Proceedings of naacL-HLT}}.
\newblock


\bibitem[Li et~al\mbox{.}(2022)]%
        {li2022learning}
\bibfield{author}{\bibinfo{person}{Haoyang Li}, \bibinfo{person}{Ziwei Zhang}, \bibinfo{person}{Xin Wang}, {and} \bibinfo{person}{Wenwu Zhu}.} \bibinfo{year}{2022}\natexlab{}.
\newblock \showarticletitle{Learning invariant graph representations for out-of-distribution generalization}.
\newblock \bibinfo{journal}{\emph{Advances in Neural Information Processing Systems}} (\bibinfo{year}{2022}), \bibinfo{pages}{11828--11841}.
\newblock


\bibitem[Li et~al\mbox{.}(2025)]%
        {li2025generating}
\bibfield{author}{\bibinfo{person}{Jin Li}, \bibinfo{person}{Shoujin Wang}, \bibinfo{person}{Qi Zhang}, \bibinfo{person}{Shui Yu}, {and} \bibinfo{person}{Fang Chen}.} \bibinfo{year}{2025}\natexlab{}.
\newblock \showarticletitle{Generating with fairness: A modality-diffused counterfactual framework for incomplete multimodal recommendations}. In \bibinfo{booktitle}{\emph{Proceedings of the ACM on Web Conference 2025}}. \bibinfo{pages}{2787--2798}.
\newblock


\bibitem[Li et~al\mbox{.}(2024)]%
        {ADRL2024MM}
\bibfield{author}{\bibinfo{person}{Zhenyang Li}, \bibinfo{person}{Fan Liu}, \bibinfo{person}{Yinwei Wei}, \bibinfo{person}{Zhiyong Cheng}, \bibinfo{person}{Liqiang Nie}, {and} \bibinfo{person}{Mohan Kankanhalli}.} \bibinfo{year}{2024}\natexlab{}.
\newblock \showarticletitle{Attribute-driven Disentangled Representation Learning for Multimodal Recommendation}. In \bibinfo{booktitle}{\emph{Proceedings of the 32nd ACM International Conference on Multimedia}}. \bibinfo{pages}{9660–9669}.
\newblock


\bibitem[Lin et~al\mbox{.}(2024)]%
        {lin2024gume}
\bibfield{author}{\bibinfo{person}{Guojiao Lin}, \bibinfo{person}{Meng Zhen}, \bibinfo{person}{Dongjie Wang}, \bibinfo{person}{Qingqing Long}, \bibinfo{person}{Yuanchun Zhou}, {and} \bibinfo{person}{Meng Xiao}.} \bibinfo{year}{2024}\natexlab{}.
\newblock \showarticletitle{GUME: Graphs and User Modalities Enhancement for Long-Tail Multimodal Recommendation}. In \bibinfo{booktitle}{\emph{Proceedings of the 33rd ACM International Conference on Information and Knowledge Management}}. \bibinfo{pages}{1400--1409}.
\newblock


\bibitem[Lin et~al\mbox{.}(2025)]%
        {lin2025contrastive}
\bibfield{author}{\bibinfo{person}{Xixun Lin}, \bibinfo{person}{Rui Liu}, \bibinfo{person}{Yanan Cao}, \bibinfo{person}{Lixin Zou}, \bibinfo{person}{Qian Li}, \bibinfo{person}{Yongxuan Wu}, \bibinfo{person}{Yang Liu}, \bibinfo{person}{Dawei Yin}, {and} \bibinfo{person}{Guandong Xu}.} \bibinfo{year}{2025}\natexlab{}.
\newblock \showarticletitle{Contrastive Modality-Disentangled Learning for Multimodal Recommendation}.
\newblock \bibinfo{journal}{\emph{ACM Transactions on Information Systems}} (\bibinfo{year}{2025}).
\newblock


\bibitem[Lin et~al\mbox{.}(2023)]%
        {lin2023contrastive}
\bibfield{author}{\bibinfo{person}{Zhenghong Lin}, \bibinfo{person}{Yanchao Tan}, \bibinfo{person}{Yunfei Zhan}, \bibinfo{person}{Weiming Liu}, \bibinfo{person}{Fan Wang}, \bibinfo{person}{Chaochao Chen}, \bibinfo{person}{Shiping Wang}, {and} \bibinfo{person}{Carl Yang}.} \bibinfo{year}{2023}\natexlab{}.
\newblock \showarticletitle{Contrastive intra-and inter-modality generation for enhancing incomplete multimedia recommendation}. In \bibinfo{booktitle}{\emph{Proceedings of the 31st ACM International Conference on Multimedia}}. \bibinfo{pages}{6234--6242}.
\newblock


\bibitem[Liu et~al\mbox{.}(2023b)]%
        {liu2023debiased}
\bibfield{author}{\bibinfo{person}{Dugang Liu}, \bibinfo{person}{Pengxiang Cheng}, \bibinfo{person}{Hong Zhu}, \bibinfo{person}{Zhenhua Dong}, \bibinfo{person}{Xiuqiang He}, \bibinfo{person}{Weike Pan}, {and} \bibinfo{person}{Zhong Ming}.} \bibinfo{year}{2023}\natexlab{b}.
\newblock \showarticletitle{Debiased representation learning in recommendation via information bottleneck}.
\newblock \bibinfo{journal}{\emph{ACM Transactions on Recommender Systems}} (\bibinfo{year}{2023}), \bibinfo{pages}{1--27}.
\newblock


\bibitem[Liu et~al\mbox{.}(2022)]%
        {liu2022disentangled}
\bibfield{author}{\bibinfo{person}{Fan Liu}, \bibinfo{person}{Huilin Chen}, \bibinfo{person}{Zhiyong Cheng}, \bibinfo{person}{Anan Liu}, \bibinfo{person}{Liqiang Nie}, {and} \bibinfo{person}{Mohan Kankanhalli}.} \bibinfo{year}{2022}\natexlab{}.
\newblock \showarticletitle{Disentangled multimodal representation learning for recommendation}.
\newblock \bibinfo{journal}{\emph{IEEE Transactions on Multimedia}}  \bibinfo{volume}{25} (\bibinfo{year}{2022}), \bibinfo{pages}{7149--7159}.
\newblock


\bibitem[Liu et~al\mbox{.}(2023a)]%
        {liu2023semantic}
\bibfield{author}{\bibinfo{person}{Fan Liu}, \bibinfo{person}{Huilin Chen}, \bibinfo{person}{Zhiyong Cheng}, \bibinfo{person}{Liqiang Nie}, {and} \bibinfo{person}{Mohan Kankanhalli}.} \bibinfo{year}{2023}\natexlab{a}.
\newblock \showarticletitle{Semantic-guided feature distillation for multimodal recommendation}. In \bibinfo{booktitle}{\emph{Proceedings of the 31st ACM International Conference on Multimedia}}. \bibinfo{pages}{6567--6575}.
\newblock


\bibitem[Liu et~al\mbox{.}(2021)]%
        {Liu2021IMP_GCN}
\bibfield{author}{\bibinfo{person}{Fan Liu}, \bibinfo{person}{Zhiyong Cheng}, \bibinfo{person}{Lei Zhu}, \bibinfo{person}{Zan Gao}, {and} \bibinfo{person}{Liqiang Nie}.} \bibinfo{year}{2021}\natexlab{}.
\newblock \showarticletitle{Interest-aware message-passing GCN for recommendation}. In \bibinfo{booktitle}{\emph{Proceedings of the web conference 2021}}. \bibinfo{pages}{1296--1305}.
\newblock


\bibitem[Liu et~al\mbox{.}(2025)]%
        {Liu20205SALLM}
\bibfield{author}{\bibinfo{person}{Fan Liu}, \bibinfo{person}{Yaqi Liu}, \bibinfo{person}{Huilin Chen}, \bibinfo{person}{Zhiyong Cheng}, \bibinfo{person}{Liqiang Nie}, {and} \bibinfo{person}{Mohan Kankanhalli}.} \bibinfo{year}{2025}\natexlab{}.
\newblock \showarticletitle{Understanding before recommendation: Semantic aspect-aware review exploitation via large language models}.
\newblock \bibinfo{journal}{\emph{ACM Transactions on Information Systems}} (\bibinfo{year}{2025}), \bibinfo{pages}{1--26}.
\newblock


\bibitem[Ma et~al\mbox{.}(2024)]%
        {ma2024multimodal}
\bibfield{author}{\bibinfo{person}{Haokai Ma}, \bibinfo{person}{Yimeng Yang}, \bibinfo{person}{Lei Meng}, \bibinfo{person}{Ruobing Xie}, {and} \bibinfo{person}{Xiangxu Meng}.} \bibinfo{year}{2024}\natexlab{}.
\newblock \showarticletitle{Multimodal conditioned diffusion model for recommendation}. In \bibinfo{booktitle}{\emph{Companion Proceedings of the ACM Web Conference 2024}}. \bibinfo{pages}{1733--1740}.
\newblock


\bibitem[Ma et~al\mbox{.}(2021)]%
        {ma2021smil}
\bibfield{author}{\bibinfo{person}{Mengmeng Ma}, \bibinfo{person}{Jian Ren}, \bibinfo{person}{Long Zhao}, \bibinfo{person}{Sergey Tulyakov}, \bibinfo{person}{Cathy Wu}, {and} \bibinfo{person}{Xi Peng}.} \bibinfo{year}{2021}\natexlab{}.
\newblock \showarticletitle{Smil: Multimodal learning with severely missing modality}. In \bibinfo{booktitle}{\emph{Proceedings of the AAAI conference on artificial intelligence}}. \bibinfo{pages}{2302--2310}.
\newblock


\bibitem[Malitesta et~al\mbox{.}(2024a)]%
        {malitesta2024we}
\bibfield{author}{\bibinfo{person}{Daniele Malitesta}, \bibinfo{person}{Emanuele Rossi}, \bibinfo{person}{Claudio Pomo}, \bibinfo{person}{Tommaso Di~Noia}, {and} \bibinfo{person}{Fragkiskos~D Malliaros}.} \bibinfo{year}{2024}\natexlab{a}.
\newblock \showarticletitle{Do We Really Need to Drop Items with Missing Modalities in Multimodal Recommendation?}. In \bibinfo{booktitle}{\emph{Proceedings of the 33rd ACM International Conference on Information and Knowledge Management}}. \bibinfo{pages}{3943--3948}.
\newblock


\bibitem[Malitesta et~al\mbox{.}(2024b)]%
        {malitesta2024dealing}
\bibfield{author}{\bibinfo{person}{Daniele Malitesta}, \bibinfo{person}{Emanuele Rossi}, \bibinfo{person}{Claudio Pomo}, \bibinfo{person}{Fragkiskos~D Malliaros}, {and} \bibinfo{person}{Tommaso Di~Noia}.} \bibinfo{year}{2024}\natexlab{b}.
\newblock \showarticletitle{Dealing with Missing Modalities in Multimodal Recommendation: a Feature Propagation-based Approach}.
\newblock \bibinfo{journal}{\emph{arXiv preprint arXiv:2403.19841}} (\bibinfo{year}{2024}).
\newblock


\bibitem[Ong and Khong(2025)]%
        {ong2025spectrum}
\bibfield{author}{\bibinfo{person}{Rongqing~Kenneth Ong} {and} \bibinfo{person}{Andy~WH Khong}.} \bibinfo{year}{2025}\natexlab{}.
\newblock \showarticletitle{Spectrum-based Modality Representation Fusion Graph Convolutional Network for Multimodal Recommendation}. In \bibinfo{booktitle}{\emph{Proceedings of the Eighteenth ACM International Conference on Web Search and Data Mining}}. \bibinfo{pages}{773--781}.
\newblock


\bibitem[Oord et~al\mbox{.}(2018)]%
        {oord2018representation}
\bibfield{author}{\bibinfo{person}{Aaron van~den Oord}, \bibinfo{person}{Yazhe Li}, {and} \bibinfo{person}{Oriol Vinyals}.} \bibinfo{year}{2018}\natexlab{}.
\newblock \showarticletitle{Representation learning with contrastive predictive coding}.
\newblock \bibinfo{journal}{\emph{arXiv preprint arXiv:1807.03748}} (\bibinfo{year}{2018}).
\newblock


\bibitem[Parthasarathy and Sundaram(2020)]%
        {parthasarathy2020training}
\bibfield{author}{\bibinfo{person}{Srinivas Parthasarathy} {and} \bibinfo{person}{Shiva Sundaram}.} \bibinfo{year}{2020}\natexlab{}.
\newblock \showarticletitle{Training strategies to handle missing modalities for audio-visual expression recognition}. In \bibinfo{booktitle}{\emph{Companion Publication of the 2020 International Conference on Multimodal Interaction}}. \bibinfo{pages}{400--404}.
\newblock


\bibitem[Rendle et~al\mbox{.}(2009)]%
        {rendle2009bpr}
\bibfield{author}{\bibinfo{person}{Steffen Rendle}, \bibinfo{person}{Christoph Freudenthaler}, \bibinfo{person}{Zeno Gantner}, {and} \bibinfo{person}{Lars Schmidt-Thieme}.} \bibinfo{year}{2009}\natexlab{}.
\newblock \showarticletitle{BPR: Bayesian personalized ranking from implicit feedback}. In \bibinfo{booktitle}{\emph{Proceedings of the Twenty-Fifth Conference on Uncertainty in Artificial Intelligence}}. \bibinfo{pages}{452--461}.
\newblock


\bibitem[Rojas-Carulla et~al\mbox{.}(2018)]%
        {rojas2018invariant}
\bibfield{author}{\bibinfo{person}{Mateo Rojas-Carulla}, \bibinfo{person}{Bernhard Sch{\"o}lkopf}, \bibinfo{person}{Richard Turner}, {and} \bibinfo{person}{Jonas Peters}.} \bibinfo{year}{2018}\natexlab{}.
\newblock \showarticletitle{Invariant models for causal transfer learning}.
\newblock \bibinfo{journal}{\emph{Journal of Machine Learning Research}} (\bibinfo{year}{2018}), \bibinfo{pages}{1--34}.
\newblock


\bibitem[Saeed et~al\mbox{.}(2024)]%
        {saeed2024modality}
\bibfield{author}{\bibinfo{person}{Muhammad~Saad Saeed}, \bibinfo{person}{Shah Nawaz}, \bibinfo{person}{Muhammad~Zaigham Zaheer}, \bibinfo{person}{Muhammad~Haris Khan}, \bibinfo{person}{Karthik Nandakumar}, \bibinfo{person}{Muhammad~Haroon Yousaf}, \bibinfo{person}{Hassan Sajjad}, \bibinfo{person}{Tom De~Schepper}, {and} \bibinfo{person}{Markus Schedl}.} \bibinfo{year}{2024}\natexlab{}.
\newblock \showarticletitle{Modality Invariant Multimodal Learning to Handle Missing Modalities: A Single-Branch Approach}.
\newblock \bibinfo{journal}{\emph{arXiv preprint arXiv:2408.07445}} (\bibinfo{year}{2024}).
\newblock


\bibitem[Simonyan and Zisserman(2015)]%
        {Simonyan2015VeryDC}
\bibfield{author}{\bibinfo{person}{K Simonyan} {and} \bibinfo{person}{A Zisserman}.} \bibinfo{year}{2015}\natexlab{}.
\newblock \showarticletitle{Very deep convolutional networks for large-scale image recognition}. In \bibinfo{booktitle}{\emph{3rd International Conference on Learning Representations}}.
\newblock


\bibitem[Sun et~al\mbox{.}(2019)]%
        {sun2019infograph}
\bibfield{author}{\bibinfo{person}{Fan-Yun Sun}, \bibinfo{person}{Jordan Hoffmann}, \bibinfo{person}{Vikas Verma}, {and} \bibinfo{person}{Jian Tang}.} \bibinfo{year}{2019}\natexlab{}.
\newblock \showarticletitle{Infograph: Unsupervised and semi-supervised graph-level representation learning via mutual information maximization}.
\newblock \bibinfo{journal}{\emph{arXiv preprint arXiv:1908.01000}} (\bibinfo{year}{2019}).
\newblock


\bibitem[Tao et~al\mbox{.}(2022)]%
        {tao2022self}
\bibfield{author}{\bibinfo{person}{Zhulin Tao}, \bibinfo{person}{Xiaohao Liu}, \bibinfo{person}{Yewei Xia}, \bibinfo{person}{Xiang Wang}, \bibinfo{person}{Lifang Yang}, \bibinfo{person}{Xianglin Huang}, {and} \bibinfo{person}{Tat-Seng Chua}.} \bibinfo{year}{2022}\natexlab{}.
\newblock \showarticletitle{Self-supervised learning for multimedia recommendation}.
\newblock \bibinfo{journal}{\emph{IEEE Transactions on Multimedia}} (\bibinfo{year}{2022}), \bibinfo{pages}{5107--5116}.
\newblock


\bibitem[Tishby et~al\mbox{.}(2000)]%
        {tishby2000information}
\bibfield{author}{\bibinfo{person}{Naftali Tishby}, \bibinfo{person}{Fernando~C Pereira}, {and} \bibinfo{person}{William Bialek}.} \bibinfo{year}{2000}\natexlab{}.
\newblock \showarticletitle{The information bottleneck method}.
\newblock \bibinfo{journal}{\emph{arXiv preprint physics/0004057}} (\bibinfo{year}{2000}).
\newblock


\bibitem[Tishby and Zaslavsky(2015)]%
        {tishby2015deep}
\bibfield{author}{\bibinfo{person}{Naftali Tishby} {and} \bibinfo{person}{Noga Zaslavsky}.} \bibinfo{year}{2015}\natexlab{}.
\newblock \showarticletitle{Deep learning and the information bottleneck principle}. In \bibinfo{booktitle}{\emph{2015 IEEE information theory workshop (itw)}}. IEEE, \bibinfo{pages}{1--5}.
\newblock


\bibitem[Wan et~al\mbox{.}(2021)]%
        {wan2021multi}
\bibfield{author}{\bibinfo{person}{Zhibin Wan}, \bibinfo{person}{Changqing Zhang}, \bibinfo{person}{Pengfei Zhu}, {and} \bibinfo{person}{Qinghua Hu}.} \bibinfo{year}{2021}\natexlab{}.
\newblock \showarticletitle{Multi-view information-bottleneck representation learning}. In \bibinfo{booktitle}{\emph{Proceedings of the AAAI conference on artificial intelligence}}. \bibinfo{pages}{10085--10092}.
\newblock


\bibitem[Wang et~al\mbox{.}(2018)]%
        {wang2018lrmm}
\bibfield{author}{\bibinfo{person}{Cheng Wang}, \bibinfo{person}{Mathias Niepert}, {and} \bibinfo{person}{Hui Li}.} \bibinfo{year}{2018}\natexlab{}.
\newblock \showarticletitle{LRMM: Learning to recommend with missing modalities}.
\newblock \bibinfo{journal}{\emph{arXiv preprint arXiv:1808.06791}} (\bibinfo{year}{2018}).
\newblock


\bibitem[Wang et~al\mbox{.}(2023a)]%
        {wang2023multi}
\bibfield{author}{\bibinfo{person}{Hu Wang}, \bibinfo{person}{Yuanhong Chen}, \bibinfo{person}{Congbo Ma}, \bibinfo{person}{Jodie Avery}, \bibinfo{person}{Louise Hull}, {and} \bibinfo{person}{Gustavo Carneiro}.} \bibinfo{year}{2023}\natexlab{a}.
\newblock \showarticletitle{Multi-modal learning with missing modality via shared-specific feature modelling}. In \bibinfo{booktitle}{\emph{Proceedings of the IEEE/CVF Conference on Computer Vision and Pattern Recognition}}. \bibinfo{pages}{15878--15887}.
\newblock


\bibitem[Wang et~al\mbox{.}(2023c)]%
        {wang2023empower}
\bibfield{author}{\bibinfo{person}{Jihong Wang}, \bibinfo{person}{Minnan Luo}, \bibinfo{person}{Jundong Li}, \bibinfo{person}{Yun Lin}, \bibinfo{person}{Yushun Dong}, \bibinfo{person}{Jin~Song Dong}, {and} \bibinfo{person}{Qinghua Zheng}.} \bibinfo{year}{2023}\natexlab{c}.
\newblock \showarticletitle{Empower post-hoc graph explanations with information bottleneck: A pre-training and fine-tuning perspective}. In \bibinfo{booktitle}{\emph{Proceedings of the 29th ACM SIGKDD Conference on Knowledge Discovery and Data Mining}}. \bibinfo{pages}{2349--2360}.
\newblock


\bibitem[Wang et~al\mbox{.}(2021)]%
        {wang2021dualgnn}
\bibfield{author}{\bibinfo{person}{Qifan Wang}, \bibinfo{person}{Yinwei Wei}, \bibinfo{person}{Jianhua Yin}, \bibinfo{person}{Jianlong Wu}, \bibinfo{person}{Xuemeng Song}, {and} \bibinfo{person}{Liqiang Nie}.} \bibinfo{year}{2021}\natexlab{}.
\newblock \showarticletitle{Dualgnn: Dual graph neural network for multimedia recommendation}.
\newblock \bibinfo{journal}{\emph{IEEE Transactions on Multimedia}}  \bibinfo{volume}{25} (\bibinfo{year}{2021}), \bibinfo{pages}{1074--1084}.
\newblock


\bibitem[Wang et~al\mbox{.}(2020)]%
        {wang2020multimodal}
\bibfield{author}{\bibinfo{person}{Qi Wang}, \bibinfo{person}{Liang Zhan}, \bibinfo{person}{Paul Thompson}, {and} \bibinfo{person}{Jiayu Zhou}.} \bibinfo{year}{2020}\natexlab{}.
\newblock \showarticletitle{Multimodal learning with incomplete modalities by knowledge distillation}. In \bibinfo{booktitle}{\emph{Proceedings of the 26th ACM SIGKDD International Conference on Knowledge Discovery \& Data Mining}}. \bibinfo{pages}{1828--1838}.
\newblock


\bibitem[Wang et~al\mbox{.}(2023b)]%
        {wang2023generative}
\bibfield{author}{\bibinfo{person}{Wenjie Wang}, \bibinfo{person}{Xinyu Lin}, \bibinfo{person}{Fuli Feng}, \bibinfo{person}{Xiangnan He}, {and} \bibinfo{person}{Tat-Seng Chua}.} \bibinfo{year}{2023}\natexlab{b}.
\newblock \showarticletitle{Generative recommendation: Towards next-generation recommender paradigm}.
\newblock \bibinfo{journal}{\emph{arXiv preprint arXiv:2304.03516}} (\bibinfo{year}{2023}).
\newblock


\bibitem[Wang et~al\mbox{.}(2019)]%
        {wang2019ngcf}
\bibfield{author}{\bibinfo{person}{Xiang Wang}, \bibinfo{person}{Xiangnan He}, \bibinfo{person}{Meng Wang}, \bibinfo{person}{Fuli Feng}, {and} \bibinfo{person}{Tat-Seng Chua}.} \bibinfo{year}{2019}\natexlab{}.
\newblock \showarticletitle{Neural graph collaborative filtering}. In \bibinfo{booktitle}{\emph{Proceedings of the 42nd international ACM SIGIR conference on Research and development in Information Retrieval}}. \bibinfo{pages}{165--174}.
\newblock


\bibitem[Wang et~al\mbox{.}(2022)]%
        {wang2022invariant}
\bibfield{author}{\bibinfo{person}{Zimu Wang}, \bibinfo{person}{Yue He}, \bibinfo{person}{Jiashuo Liu}, \bibinfo{person}{Wenchao Zou}, \bibinfo{person}{Philip~S Yu}, {and} \bibinfo{person}{Peng Cui}.} \bibinfo{year}{2022}\natexlab{}.
\newblock \showarticletitle{Invariant preference learning for general debiasing in recommendation}. In \bibinfo{booktitle}{\emph{Proceedings of the 28th ACM SIGKDD Conference on Knowledge Discovery and Data Mining}}. \bibinfo{pages}{1969--1978}.
\newblock


\bibitem[Wei et~al\mbox{.}(2022)]%
        {wei2022contrastive}
\bibfield{author}{\bibinfo{person}{Chunyu Wei}, \bibinfo{person}{Jian Liang}, \bibinfo{person}{Di Liu}, {and} \bibinfo{person}{Fei Wang}.} \bibinfo{year}{2022}\natexlab{}.
\newblock \showarticletitle{Contrastive graph structure learning via information bottleneck for recommendation}.
\newblock \bibinfo{journal}{\emph{Advances in neural information processing systems}} (\bibinfo{year}{2022}), \bibinfo{pages}{20407--20420}.
\newblock


\bibitem[Wei et~al\mbox{.}(2023a)]%
        {wei2023multi}
\bibfield{author}{\bibinfo{person}{Wei Wei}, \bibinfo{person}{Chao Huang}, \bibinfo{person}{Lianghao Xia}, {and} \bibinfo{person}{Chuxu Zhang}.} \bibinfo{year}{2023}\natexlab{a}.
\newblock \showarticletitle{Multi-modal self-supervised learning for recommendation}. In \bibinfo{booktitle}{\emph{Proceedings of the ACM Web Conference 2023}}. \bibinfo{pages}{790--800}.
\newblock


\bibitem[Wei et~al\mbox{.}(2023b)]%
        {Wei2023LightGT}
\bibfield{author}{\bibinfo{person}{Yinwei Wei}, \bibinfo{person}{Wenqi Liu}, \bibinfo{person}{Fan Liu}, \bibinfo{person}{Xiang Wang}, \bibinfo{person}{Liqiang Nie}, {and} \bibinfo{person}{Tat-Seng Chua}.} \bibinfo{year}{2023}\natexlab{b}.
\newblock \showarticletitle{Lightgt: A light graph transformer for multimedia recommendation}. In \bibinfo{booktitle}{\emph{Proceedings of the 46th international ACM SIGIR conference on research and development in information retrieval}}. \bibinfo{pages}{1508--1517}.
\newblock


\bibitem[Wei et~al\mbox{.}(2020)]%
        {Wei2020GRCN}
\bibfield{author}{\bibinfo{person}{Yinwei Wei}, \bibinfo{person}{Xiang Wang}, \bibinfo{person}{Liqiang Nie}, \bibinfo{person}{Xiangnan He}, {and} \bibinfo{person}{Tat-Seng Chua}.} \bibinfo{year}{2020}\natexlab{}.
\newblock \showarticletitle{Graph-refined convolutional network for multimedia recommendation with implicit feedback}. In \bibinfo{booktitle}{\emph{Proceedings of the 28th ACM international conference on multimedia}}. \bibinfo{pages}{3541--3549}.
\newblock


\bibitem[Wei et~al\mbox{.}(2019)]%
        {wei2019mmgcn}
\bibfield{author}{\bibinfo{person}{Yinwei Wei}, \bibinfo{person}{Xiang Wang}, \bibinfo{person}{Liqiang Nie}, \bibinfo{person}{Xiangnan He}, \bibinfo{person}{Richang Hong}, {and} \bibinfo{person}{Tat-Seng Chua}.} \bibinfo{year}{2019}\natexlab{}.
\newblock \showarticletitle{MMGCN: Multi-modal graph convolution network for personalized recommendation of micro-video}. In \bibinfo{booktitle}{\emph{Proceedings of the 27th ACM international conference on multimedia}}. \bibinfo{pages}{1437--1445}.
\newblock


\bibitem[Woo et~al\mbox{.}(2023)]%
        {woo2023towards}
\bibfield{author}{\bibinfo{person}{Sangmin Woo}, \bibinfo{person}{Sumin Lee}, \bibinfo{person}{Yeonju Park}, \bibinfo{person}{Muhammad~Adi Nugroho}, {and} \bibinfo{person}{Changick Kim}.} \bibinfo{year}{2023}\natexlab{}.
\newblock \showarticletitle{Towards good practices for missing modality robust action recognition}. In \bibinfo{booktitle}{\emph{Proceedings of the AAAI Conference on Artificial Intelligence}}, Vol.~\bibinfo{volume}{37}. \bibinfo{pages}{2776--2784}.
\newblock


\bibitem[Wu et~al\mbox{.}(2020b)]%
        {wu2020joint}
\bibfield{author}{\bibinfo{person}{Le Wu}, \bibinfo{person}{Yonghui Yang}, \bibinfo{person}{Kun Zhang}, \bibinfo{person}{Richang Hong}, \bibinfo{person}{Yanjie Fu}, {and} \bibinfo{person}{Meng Wang}.} \bibinfo{year}{2020}\natexlab{b}.
\newblock \showarticletitle{Joint item recommendation and attribute inference: An adaptive graph convolutional network approach}. In \bibinfo{booktitle}{\emph{Proceedings of the 43rd International ACM SIGIR conference on research and development in Information Retrieval}}. \bibinfo{pages}{679--688}.
\newblock


\bibitem[Wu et~al\mbox{.}(2020a)]%
        {wu2020graph}
\bibfield{author}{\bibinfo{person}{Tailin Wu}, \bibinfo{person}{Hongyu Ren}, \bibinfo{person}{Pan Li}, {and} \bibinfo{person}{Jure Leskovec}.} \bibinfo{year}{2020}\natexlab{a}.
\newblock \showarticletitle{Graph information bottleneck}.
\newblock \bibinfo{journal}{\emph{Advances in Neural Information Processing Systems}}  \bibinfo{volume}{33} (\bibinfo{year}{2020}), \bibinfo{pages}{20437--20448}.
\newblock


\bibitem[Xu et~al\mbox{.}(2024)]%
        {xu2024mentor}
\bibfield{author}{\bibinfo{person}{Jinfeng Xu}, \bibinfo{person}{Zheyu Chen}, \bibinfo{person}{Shuo Yang}, \bibinfo{person}{Jinze Li}, \bibinfo{person}{Hewei Wang}, {and} \bibinfo{person}{Edith C-H Ngai}.} \bibinfo{year}{2024}\natexlab{}.
\newblock \showarticletitle{Mentor: multi-level self-supervised learning for multimodal recommendation}.
\newblock \bibinfo{journal}{\emph{arXiv preprint arXiv:2402.19407}} (\bibinfo{year}{2024}).
\newblock


\bibitem[Yang et~al\mbox{.}(2025)]%
        {yang2025less}
\bibfield{author}{\bibinfo{person}{Yonghui Yang}, \bibinfo{person}{Le Wu}, \bibinfo{person}{Zhuangzhuang He}, \bibinfo{person}{Zhengwei Wu}, \bibinfo{person}{Richang Hong}, {and} \bibinfo{person}{Meng Wang}.} \bibinfo{year}{2025}\natexlab{}.
\newblock \showarticletitle{Less is More: Information Bottleneck Denoised Multimedia Recommendation}.
\newblock \bibinfo{journal}{\emph{arXiv preprint arXiv:2501.12175}} (\bibinfo{year}{2025}).
\newblock


\bibitem[Yang et~al\mbox{.}(2024)]%
        {yang2024graph}
\bibfield{author}{\bibinfo{person}{Yonghui Yang}, \bibinfo{person}{Le Wu}, \bibinfo{person}{Zihan Wang}, \bibinfo{person}{Zhuangzhuang He}, \bibinfo{person}{Richang Hong}, {and} \bibinfo{person}{Meng Wang}.} \bibinfo{year}{2024}\natexlab{}.
\newblock \showarticletitle{Graph bottlenecked social recommendation}. In \bibinfo{booktitle}{\emph{Proceedings of the 30th ACM SIGKDD Conference on Knowledge Discovery and Data Mining}}. \bibinfo{pages}{3853--3862}.
\newblock


\bibitem[You et~al\mbox{.}(2020)]%
        {you2020graph}
\bibfield{author}{\bibinfo{person}{Yuning You}, \bibinfo{person}{Tianlong Chen}, \bibinfo{person}{Yongduo Sui}, \bibinfo{person}{Ting Chen}, \bibinfo{person}{Zhangyang Wang}, {and} \bibinfo{person}{Yang Shen}.} \bibinfo{year}{2020}\natexlab{}.
\newblock \showarticletitle{Graph contrastive learning with augmentations}.
\newblock \bibinfo{journal}{\emph{Advances in neural information processing systems}}  \bibinfo{volume}{33} (\bibinfo{year}{2020}), \bibinfo{pages}{5812--5823}.
\newblock


\bibitem[Yu et~al\mbox{.}(2023)]%
        {yu2023multi}
\bibfield{author}{\bibinfo{person}{Penghang Yu}, \bibinfo{person}{Zhiyi Tan}, \bibinfo{person}{Guanming Lu}, {and} \bibinfo{person}{Bing-Kun Bao}.} \bibinfo{year}{2023}\natexlab{}.
\newblock \showarticletitle{Multi-view graph convolutional network for multimedia recommendation}. In \bibinfo{booktitle}{\emph{Proceedings of the 31st ACM international conference on multimedia}}. \bibinfo{pages}{6576--6585}.
\newblock


\bibitem[Zhang et~al\mbox{.}(2023b)]%
        {zhang2023invariant}
\bibfield{author}{\bibinfo{person}{An Zhang}, \bibinfo{person}{Jingnan Zheng}, \bibinfo{person}{Xiang Wang}, \bibinfo{person}{Yancheng Yuan}, {and} \bibinfo{person}{Tat-Seng Chua}.} \bibinfo{year}{2023}\natexlab{b}.
\newblock \showarticletitle{Invariant collaborative filtering to popularity distribution shift}. In \bibinfo{booktitle}{\emph{Proceedings of the ACM Web Conference 2023}}. \bibinfo{pages}{1240--1251}.
\newblock


\bibitem[Zhang et~al\mbox{.}(2022)]%
        {zhang2022improving}
\bibfield{author}{\bibinfo{person}{Cenyuan Zhang}, \bibinfo{person}{Xiang Zhou}, \bibinfo{person}{Yixin Wan}, \bibinfo{person}{Xiaoqing Zheng}, \bibinfo{person}{Kai-Wei Chang}, {and} \bibinfo{person}{Cho-Jui Hsieh}.} \bibinfo{year}{2022}\natexlab{}.
\newblock \showarticletitle{Improving the adversarial robustness of NLP models by information bottleneck}.
\newblock \bibinfo{journal}{\emph{arXiv preprint arXiv:2206.05511}} (\bibinfo{year}{2022}).
\newblock


\bibitem[Zhang et~al\mbox{.}(2021)]%
        {zhang2021mining}
\bibfield{author}{\bibinfo{person}{Jinghao Zhang}, \bibinfo{person}{Yanqiao Zhu}, \bibinfo{person}{Qiang Liu}, \bibinfo{person}{Shu Wu}, \bibinfo{person}{Shuhui Wang}, {and} \bibinfo{person}{Liang Wang}.} \bibinfo{year}{2021}\natexlab{}.
\newblock \showarticletitle{Mining latent structures for multimedia recommendation}. In \bibinfo{booktitle}{\emph{Proceedings of the 29th ACM international conference on multimedia}}. \bibinfo{pages}{3872--3880}.
\newblock


\bibitem[Zhang et~al\mbox{.}(2023a)]%
        {zhang2023reformulating}
\bibfield{author}{\bibinfo{person}{Yang Zhang}, \bibinfo{person}{Tianhao Shi}, \bibinfo{person}{Fuli Feng}, \bibinfo{person}{Wenjie Wang}, \bibinfo{person}{Dingxian Wang}, \bibinfo{person}{Xiangnan He}, {and} \bibinfo{person}{Yongdong Zhang}.} \bibinfo{year}{2023}\natexlab{a}.
\newblock \showarticletitle{Reformulating CTR prediction: Learning invariant feature interactions for recommendation}. In \bibinfo{booktitle}{\emph{Proceedings of the 46th International ACM SIGIR Conference on Research and Development in Information Retrieval}}. \bibinfo{pages}{1386--1395}.
\newblock


\bibitem[Zhao et~al\mbox{.}(2025)]%
        {zhao2025dvib}
\bibfield{author}{\bibinfo{person}{Wenkuan Zhao}, \bibinfo{person}{Shanshan Zhong}, \bibinfo{person}{Yifan Liu}, \bibinfo{person}{Wushao Wen}, \bibinfo{person}{Jinghui Qin}, \bibinfo{person}{Mingfu Liang}, {and} \bibinfo{person}{Zhongzhan Huang}.} \bibinfo{year}{2025}\natexlab{}.
\newblock \showarticletitle{DVIB: Towards Robust Multimodal Recommender Systems via Variational Information Bottleneck Distillation}. In \bibinfo{booktitle}{\emph{Proceedings of the ACM on Web Conference 2025}}. \bibinfo{pages}{2549--2561}.
\newblock


\bibitem[Zhou et~al\mbox{.}(2023b)]%
        {zhou2023enhancing}
\bibfield{author}{\bibinfo{person}{Hongyu Zhou}, \bibinfo{person}{Xin Zhou}, \bibinfo{person}{Lingzi Zhang}, {and} \bibinfo{person}{Zhiqi Shen}.} \bibinfo{year}{2023}\natexlab{b}.
\newblock \showarticletitle{Enhancing dyadic relations with homogeneous graphs for multimodal recommendation}.
\newblock In \bibinfo{booktitle}{\emph{ECAI 2023}}. \bibinfo{publisher}{IOS Press}, \bibinfo{pages}{3123--3130}.
\newblock


\bibitem[Zhou et~al\mbox{.}(2023a)]%
        {BM32020Arxiv}
\bibfield{author}{\bibinfo{person}{Xin Zhou}, \bibinfo{person}{Hongyu Zhou}, \bibinfo{person}{Yong Liu}, \bibinfo{person}{Zhiwei Zeng}, \bibinfo{person}{Chunyan Miao}, \bibinfo{person}{Pengwei Wang}, \bibinfo{person}{Yuan You}, {and} \bibinfo{person}{Feijun Jiang}.} \bibinfo{year}{2023}\natexlab{a}.
\newblock \showarticletitle{Bootstrap latent representations for multi-modal recommendation}. In \bibinfo{booktitle}{\emph{Proceedings of the ACM web conference 2023}}. \bibinfo{pages}{845--854}.
\newblock


\end{thebibliography}

\end{document}